\documentclass[aps,twocolumn,showpacs,preprintnumbers,nofootinbib,prd,10pt,superscriptaddress]{revtex4-1}

\makeatletter
\def\l@subsubsection#1#2{}
\def\l@subsubsubsection#1#2{}
\makeatother

\usepackage{graphicx,amssymb,amsmath,amsthm,amsfonts,epsfig,epsf,fixmath}
\usepackage[usenames]{color}
\usepackage{epstopdf}

\usepackage{comment}
\usepackage{tabularx}
\usepackage{bm}
\usepackage{dcolumn}
\usepackage{latexsym}
\usepackage{rotating}
\usepackage{longtable}

\usepackage{enumerate}
\usepackage{tensor,multirow}
\usepackage{url}
\usepackage[dvipsnames]{xcolor}
\usepackage[unicode]{hyperref}

\hypersetup{colorlinks=true, citecolor=MidnightBlue,
            linkcolor=MidnightBlue, urlcolor=MidnightBlue, linktocpage=true}

\newcommand{\tn}{\textnormal}

\newcommand{\GSSI}{Gran Sasso Science Institute (GSSI), I-67100 L’Aquila, Italy}
\newcommand{\GranSasso}{INFN, Laboratori Nazionali del Gran Sasso, I-67100 Assergi, Italy}

\begin{document}

\title{The multipolar structure of rotating boson stars}

\author{Massimo Vaglio}
\affiliation{Dipartimento di Fisica, ``Sapienza'' Universit\`a di Roma, Piazzale
Aldo Moro 5, 00185, Roma, Italy}
\address{INFN, Sezione di Roma, Piazzale Aldo Moro 2, 00185, Roma, Italy}
 \author{Costantino Pacilio}
\affiliation{Dipartimento di Fisica, ``Sapienza'' Universit\`a di Roma, Piazzale
Aldo Moro 5, 00185, Roma, Italy}
\address{INFN, Sezione di Roma, Piazzale Aldo Moro 2, 00185, Roma, Italy}
 \author{Andrea Maselli}
\address{\GSSI}
\address{\GranSasso}
\author{Paolo Pani}
\affiliation{Dipartimento di Fisica, ``Sapienza'' Universit\`a di Roma, Piazzale
Aldo Moro 5, 00185, Roma, Italy}
\address{INFN, Sezione di Roma, Piazzale Aldo Moro 2, 00185, Roma, Italy}

\date{\today}
\begin{abstract}
The relativistic multipole moments provide a key ingredient to characterize 
the gravitational field around compact astrophysical objects. 
They play a crucial role in the description of the orbital evolution of 
coalescing binary systems and encode valuable information on the nature 
of the binary’s components, which leaves a measurable imprint in their 
gravitational-wave emission. We present a new study on the multipolar structure 
of a class of arbitrarily spinning boson stars with quartic self-interactions 
in the large coupling limit, where these solutions are expected to be stable. 
Our results strengthen and extend previous numerical analyses, showing that even 
for the most compact configurations the multipolar structure deviates 
significantly from that of a Kerr black hole.
We provide accurate data for the multipole moments as functions of the object's mass and spin, which can be directly 
used to construct inspiral waveform approximants and to perform parameter estimations
and searches for boson star binaries.
\end{abstract}

\maketitle

\section{Introduction}

The advent of gravitational-wave~(GW) astronomy has opened 
new opportunities for tests of fundamental physics~\cite{Barack:2018yly}. 
A cornerstone of this program is to use GW data to probe 
the nature of compact objects~\cite{Cardoso:2019rvt,LIGOScientific:2021sio,Maggio:2021ans}, 
and in particular to explore the possibility that 
astrophysical compact sources other than black holes~(BH) 
and neutron stars can exist in the Universe.
These hypothetical objects can provide a new portal to test 
a variety of particle and high-energy physics models~\cite{Giudice:2016zpa,pacilio_gravitational-wave_2020} 
and could be an exotic explanation~\cite{Bustillo:2020syj} for the LIGO/Virgo ``mass-gap events'' 
(e.g. GW190814~\cite{LIGOScientific:2020ufj} and 
GW190521~\cite{LIGOScientific:2020iuh,LIGOScientific:2020ufj})
which do not fit naturally within the standard astrophysical 
formation scenarios for BHs and neutron stars.

Among the plethora of exotic compact objects~\cite{Cardoso:2019rvt}, 
boson stars~(BSs) stand out as one of the best motivated models 
arising from a concrete field theory. BSs are self-gravitating 
solitons, composed of either scalar~\cite{Kaup:1968zz,Ruffini:1969qy,colpi_boson_1986} 
or vector~\cite{Brito:2015pxa}, massive complex fields, 
minimally coupled to Einstein's gravity 
(see~\cite{Jetzer:1991jr,liebling_dynamical_2017} for some reviews).
At variance with other models, for BSs the whole dynamics (including 
BS mergers~\cite{Palenzuela:2007dm,Palenzuela:2017kcg,Bezares:2018qwa,Bustillo:2020syj,Bezares:2022obu} and nonlinear stability 
analysis~\cite{Sanchis-Gual:2019ljs,DiGiovanni:2020ror,siemonsen_stability_2021}) 
and phenomenology can be studied from first principles. They are 
therefore a natural target for GW searches.

Deviations in the GW inspiral signals with respect to the case of BH and neutron star
binaries can be traced back to the so-called finite-size 
effects, which encode the properties of the object's internal structure.
In a post-Newtonian expansion of Einstein's field equations for a binary system, 
the leading-order effect depending on the internal structure of the binary 
components is the spin-induced mass quadrupole moment, $M_2$~\cite{PoissonWill}. 
According to General Relativity, if the object is a stationary BH, it must be axisymmetric 
and described by the Kerr solution. Due to the symmetries of the latter, 
the multipolar structure of a BH in General Relativity is encoded in a closed-form, elegant, 
relation~\cite{Hansen:1974zz} 
\begin{equation}
 {M}_\ell^{\rm BH}+{ i }  {S}_\ell^{\rm BH}  
 ={M}^{\ell+1}\left({ i } \chi\right)^\ell\,, \label{nohair}
\end{equation}
where ${M}_\ell$ (${S}_\ell$) are the Geroch-Hansen mass (current) 
multipole moments~\cite{Geroch:1970cd,Hansen:1974zz}, ${M}\equiv M_0$ is the mass, 
$J \equiv {S}_1$ is  the angular momentum, and $\chi\equiv{J  }/{{M}^2}$ 
is the dimensionless spin\footnote{For a generic spacetime the multipole 
moments of order $\ell$ are rank-$\ell$ tensors, ${M}_{\ell m}$ and ${S}_{\ell m}$, 
which reduce to scalar quantities, ${M}_\ell$ and ${S}_\ell$, in the axisymmetric case, 
see e.g. Refs.~\cite{Bianchi:2020bxa,Bianchi:2020miz} for the general definitions. 
In this paper we shall only focus on axisymmetric and equatorial symmetric spacetimes 
and therefore we shall only deal with scalar quantities with the same symmetries 
of Kerr's (see~\cite{Fransen:2022jtw,Loutrel:2022ant} for a recent work in 
which the equatorial and axial symmetries are relaxed).}.
Introducing the dimensionless quantities ${\bar{M}}_\ell \equiv{ 
M}_\ell/{ M}^{\ell{+}1}$ and ${\bar S}_\ell \equiv{ S}_\ell/{M}^{\ell{+}1}$, 
the only nonvanishing moments of the Kerr spacetime are
\begin{equation}
 {\bar{M}}_{2n}^{\rm BH}   = (-1)^n \chi^{2n} \quad, \quad
 {\bar{S}}_{2n{+}1}^{\rm BH} = (-1)^n \chi^{2n{+}1} \label{momKerr}
\end{equation}
for $n=0,1,2,...$. Besides the fact that the entire multipolar structure 
is completely determined only by the BH mass and spin, having ${M}_\ell=0$ 
(${ S}_\ell=0$) when $\ell$ is odd (even) is a consequence of the equatorial 
symmetry of the Kerr metric, whereas the fact that all nonvanishing 
$\ell$-th multipoles (with $\ell\geq2$) are proportional to $\chi^\ell$ 
is a peculiarity of the Kerr metric.

Any deviation from the above multipolar structure would imply that 
the underlying spacetime is not described by the Kerr solution. 
Therefore, measuring any multipole moment of a compact object in addition to the mass and spin would provide a null-hypothesis tests of the Kerr metric~\cite{Psaltis:2008bb,Gair:2012nm,Yunes:2013dva,Berti:2015itd,Cardoso:2016ryw,Barack:2018yly,Cardoso:2019rvt}.

Going beyond null-hypothesis tests (e.g. if one wishes to perform model selection between the Kerr hypothesis and a more exotic model) requires computing the multipolar structure of alternative objects.
In particular, the multipolar structure of BSs differs from that of a BH and depends on the underlying scalar self-interactions~\cite{ryan_spinning_1997,pacilio_gravitational-wave_2020}, similarly to the case of neutron stars where the multipole moments depend on the underlying equation of state.

The multipolar structure of BSs with quartic scalar interactions was computed in a seminal paper by Ryan~\cite{ryan_spinning_1997} by using a perfect-fluid approximation scheme valid in the large self-coupling regime and implementing an iterative method to solve for Einstein's equations in the stationary and axisymmetric case.
The scope of this work is to extend Ryan's analysis in order to accurately compute the leading-order moments (the mass quadrupole and the current octupole) in the entire parameter space of the model, and to provide accurate data, useful 
to build waveform templates~\cite{Barack:2006pq,Krishnendu:2017shb,Krishnendu:2019ebd} for actual searches and parameter estimation.
Henceforth we adopt $G=c=1$ units.

\section{\label{sec:level1}Theoretical setup}

We consider stationary axisymmetric BSs as solutions of the
Einstein--Klein-Gordon equations for a complex, massive, 
self-interacting scalar field minimally coupled to the 
gravitational sector~\cite{liebling_dynamical_2017}.
The Lagrangian governing the field dynamics reads 
\begin{equation}
{\cal L}_\phi=-\frac{1}{2}g^{\mu\nu}\phi^*_{,\mu}\phi_{,\nu}
-\frac{1}{2}V(|\phi|^2)\ , \label{lagrangian}
\end{equation}
where $V(|\phi|^2)$ is the scalar potential, which includes 
the mass term as well as self-interactions determining the object multipole moments.
Varying the total action
\begin{equation}
{\cal S}=\int d^4 x\sqrt{-g}
\left[\frac{R}{16\pi}-{\cal L}_\phi\right]\ ,
\end{equation} 
we obtain the field's equations
\begin{subequations}
\begin{align}
G_{\mu\nu}&=8\pi T_{\mu\nu}\,,\\
\frac{1}{\sqrt{-g}}(\sqrt{-g}g^{\mu\nu}\phi_{,\mu})_{,\nu}&=
\frac{dV}{d|\phi|^2}\phi\,,
\end{align}
\label{fieldequations}
\end{subequations}
where $g$ is the metric determinant and 
$T_{\mu\nu}$ is the canonical 
stress-energy tensor
\begin{equation*}
T_{\mu\nu}=\frac{1}{2}(\phi^*_{,\mu}\phi_{,\nu}+\phi_{,\mu}\phi^*_{,\nu})-
\frac{1}{2}g_{\mu\nu}\left[g^{\alpha\beta}\phi^*_{,\alpha}\phi_{,\beta}+V(|\phi|^2)\right]\,.
\end{equation*}

We look for stationary and axisymmetric solutions of 
Eqns.~\eqref{fieldequations}. 
Using the set of coordinates 
$x^\mu=(t,r,\theta,\varphi)$ adapted to the isometry generators 
 $\bigl(\frac{\partial}{\partial{t}}, 
\frac{\partial}{\partial{\varphi}}\bigr)$, the metric and 
the stress-energy tensor of the solutions  
have no explicit dependence on $t$ and $\varphi$.
Stationarity and axisymmetry 
require the scalar field to satisfy the ansatz
\begin{equation}
\phi=\phi_0(r,\theta)e^{i(s\phi-\Omega t)}\ ,
\label{ansatz}
\end{equation} 
where the \emph{azimuthal winding number} $s$ is an integer 
related to the BS total angular momentum and $\Omega>0$ is the 
field angular frequency~\cite{herdeiro_spinning_2018}. 
We adopt quasi-isotropic coordinates for the metric of 
a stationary and axisymmetric spacetime
\begin{align}
ds^2=&-e^{\gamma+\rho}dt^2+e^{2\alpha}(dr^2+r^2d\theta^2)\nonumber\\
&+e^{\gamma-\rho}r^2\sin{\theta}^2(d\varphi-\omega dt)^2\ ,
\label{Papape}
\end{align}
where the four metric functions $(\gamma,\rho,\alpha,\omega)$, 
depend on $(r,\theta)$ only.
In this work we consider a specific family of 
\textit{massive} BSs~\cite{colpi_boson_1986,ryan_spinning_1997}, 
featuring repulsive quartic self-interactions,
\begin{equation}
V(|\phi|^2)=m^2|\phi|^2+\frac{1}{2}\lambda|\phi|^4 \ .
\end{equation}
Moreover we focus on the strong coupling limit $\lambda/m^2\gg1$, in
which the maximum mass supported by \textit{static} configurations
scales as~\cite{colpi_boson_1986}
\begin{equation}
M_\tn{\rm max}\sim 0.06\frac{\sqrt{\lambda \hbar}}{m_s^2}M_p^3\ ,\label{eq_maxmass}
\end{equation}
where $m_s=m\hbar$ is the mass of the boson and $M_p$ the 
Planck mass.
Equation~\eqref{eq_maxmass} shows that, for 
$\lambda \sim \mathcal{O}(\hbar^{-1})$ and $m_s$ in the range $1$--$100$ MeV,
stellar configurations with $M_\tn{\rm max}$ in the range $10$--$10^5~M_\odot$ are supported. This is different from the case of \textit{mini} BSs described by non-interacting scalars~\cite{Kaup:1968zz,Ruffini:1969qy}, where the same mass range requires ultralight bosons.
Moreover, large self-interactions are also expected to quench~\cite{siemonsen_stability_2021} the 
instabilities observed in numerical simulations of rotating \emph{mini} BSs~\cite{Sanchis-Gual:2019ljs,DiGiovanni:2020ror}.

\subsection{Perfect-fluid approximation in the strong-coupling limit} \label{sec:tails}

\begin{figure}[htp]
    \centering
    \includegraphics[width=7cm]{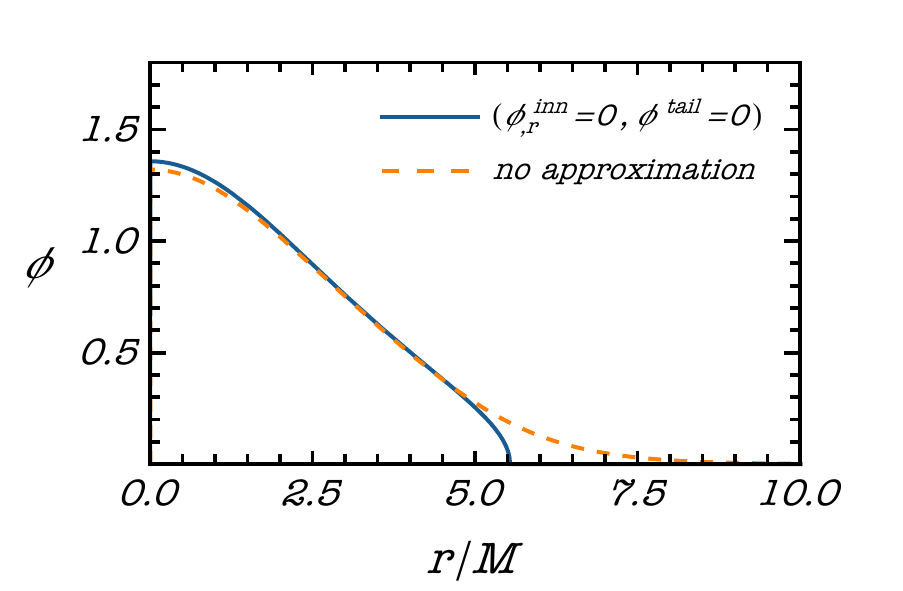}
    \caption{Scalar field profile for a spherically-symmetric 
    massive BS with $\lambda/m^2=2500$, computed (i) neglecting radial derivatives 
    of $\phi$ in the inner zone, and setting $\phi=0$ in tail 
    zone (solid line) and (ii) with no approximations on the 
    scalar field in the domain of integration (dashed line).}
    \label{approx}
\end{figure}

In the strong coupling regime the numerical integration of the 
stellar equations greatly simplifies. As discussed in 
\cite{colpi_boson_1986}, for spherically-symmetric solutions 
it is possible to identify two distinct regions in the object 
radial domain, corresponding to different 
behaviours of the field's energy density. 
At large values of $r$ the BS features a \emph{tail} region 
where $\phi$ decays exponentially as 
$\sim e^{-\sqrt{m^2-\Omega^2}r}/r$. At smaller $r$, an inner 
\emph{non-tail} region sets up, where the field has significantly 
larger amplitude and most of the object's mass is localized. 
In this zone $\phi$ varies on a very large scale, 
such that one can safely assume $\phi_{,r} \simeq 0$, while in 
the \emph{tail} region, although $\phi_{,r}$ is in general 
not negligible, the field vanishes quickly due to the the 
exponential suppression and can be set to zero. We have 
numerically confirmed the validity of these assumptions, as 
shown in Fig.~\ref{approx}, which displays the radial profile 
of $\phi$ for two spherically symmetric BSs with the same 
frequency, built considering $\phi^\tn{non-tail}_{,r} \sim 0,\ \phi^\tn{tail} \sim 0$ (solid curve) and 
without any approximation (dashed curve).

In the spinning case, a further simplification can be made.
Indeed, as noted in~\cite{ryan_spinning_1997}, the symmetry of the 
solution suggests that the field stress-energy tensor should 
vary on the same scale when the star is rotating or not, 
such that in both cases derivatives with respect to the 
radial direction and the polar angle $\theta$ can be 
neglected in the \emph{non-tail} region, while in general $\phi_{,\varphi}\neq 0$. 
Setting $\phi_{,r}$ and $\phi_{,\theta}$ to zero in the stress-energy 
tensor and using the ansatz $\eqref{ansatz}$, we can recast $T_{\mu\nu}$  
within the inner region in the following form
\begin{equation}
T_{\mu\nu}=(\epsilon+P)u_\mu u_\nu+Pg_{\mu\nu}\ ,
\end{equation}
where
\begin{subequations}
\begin{align}
&(u_t,u_r,u_\theta,u_\varphi)=A^{-1/2}(-\Omega,0,0,s)\,,\\
&A=\frac{g^{\alpha \beta}\phi^*_{,\alpha} \phi_{,\beta}}{|\phi|^2}\approx(-g^{tt}\Omega^2+2g^{t\varphi}\Omega s-g^{\varphi\varphi}s^2)\,,
\end{align}
\label{perfect}
\end{subequations}
and we identify the field's pressure and energy density 
\begin{subequations}
\begin{align}
&P=\frac{1}{2}A|\phi|^2 - \frac{1}{2}V(|\phi|^2)\,,\\
&\epsilon=\frac{1}{2}A|\phi|^2 + \frac{1}{2}V(|\phi|^2)\,.
\end{align}
\end{subequations}

In the \emph{tail} region we assume that the scalar field 
is negligible, and we set $T_{\mu\nu}=0$. Therefore, 
within the entire domain of integration, the stress-energy tensor resembles 
that of a perfect fluid.

Note that, in the inner region of a rotating massive BS, 
the energy density develops a non-trivial topology. Indeed, neglecting the radial and polar derivatives 
of $\phi$, the scalar field satisfies the
constraint equation
\begin{equation}
\Bigl(A\phi-\frac{dV}{d|\phi|^2}\phi\Bigr)=A\phi-m^2\phi-\lambda|\phi|^2\phi=0\,,
\end{equation}
whose solutions are
\begin{equation*}
  \left\{
    \begin{aligned}
       |\phi|^2&=\frac{A-m^2}{\lambda} \quad  &{\rm if} \quad  A>m^2\,, \\
       \phi&=0 \quad &{\rm if} \quad A\leq m^2\,.
    \end{aligned}
  \right.
\end{equation*}

On the other hand, outside the inner region $|\phi| \sim 0$. 
Therefore, under the above approximations, the general expression for $|\phi|^2$ reads
\begin{align}
|\phi|^2&={\rm max}[0,(-g^{tt}\Omega^2+2g^{t\varphi} \Omega s-g^{\varphi\varphi} s^2 -m^2)/\lambda]\nonumber\\
\phantom{a}\nonumber\\
&={\rm max}\left[0,\frac{1}{\lambda}\left(\frac{(\Omega-s\omega)^2}{e^{\gamma+\rho}}-\frac{e^{\rho-\gamma}s^2}{r^2\sin{\theta}^2}-m^2\right)\right]\,.
\label{FieldExpr}
\end{align}

The pressure, energy density, and four-velocity can be expressed in terms of $|\phi|^2$ as\footnote{Therefore, in this approximation the interior of the star is described by a perfect fluid with a barotropic equation of state
\begin{equation}
P(\epsilon)=\frac{m^4}{\lambda}\left(1+\sqrt{1+\frac{3\epsilon\lambda}{m^4}}\right)^2\ .
\end{equation}
}
\begin{subequations}
\begin{align}
&P=\frac{1}{4}\lambda|\phi|^4\,,\\
&\epsilon=m^2|\phi|^2+\frac{3}{4}\lambda|\phi|^4\,,\\
&(u_t,u_r,u_\theta,u_\varphi)=\frac{(-\Omega,0,0,s)}{(\lambda|\phi|^2+m^2)^\frac{1}{2}}\,.
\end{align}
\label{eq:peu}
\end{subequations}

\begin{figure}[htp]
    \centering
    \includegraphics[width=0.49\textwidth]{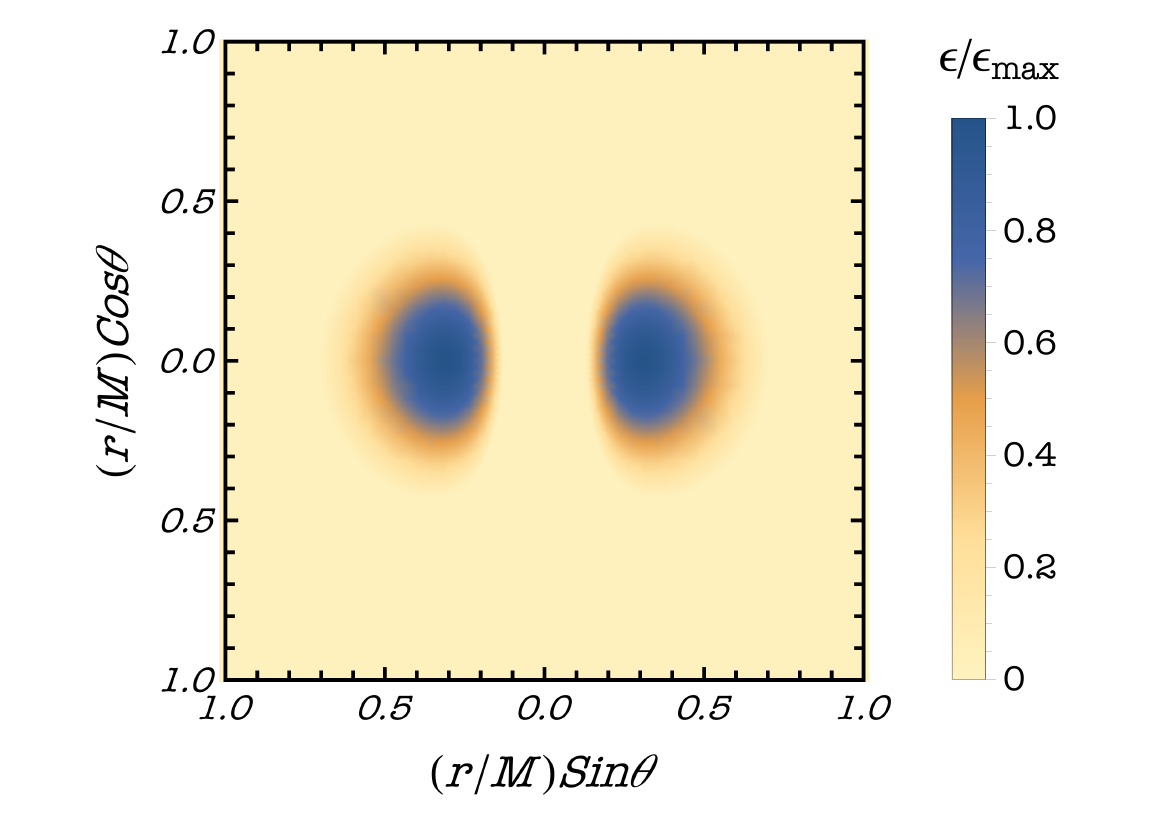}
    \caption{Vertical cross-section of a fast rotating BS with $M=0.04 M_B$ 
    and $\chi=J/M^2=1.3$. The scalar field energy density is normalized to its maximum value.}
    \label{fig:scalar_density}
\end{figure}

By combining Eqns.~\eqref{FieldExpr}-\eqref{eq:peu} one can see that the energy density of rotating BSs develops a toroidal shape, as 
evident from  Eq.~\eqref{FieldExpr} which shows that $|\phi|^2$ 
is zero near the polar axis where $\sin\theta\approx 0$. This behaviour is displayed in Fig.~\ref{fig:scalar_density} for a representative model. 
Note also that, in the absence of rotation, the torus degenerates into a spherical profile.

\subsection{Coordinate rescaling} \label{Sec:rescaling}

Our numerical analysis can be further simplified by a suitable 
change of variables which removes both $m$ and $\lambda$ from the 
field's equations.  
In geometrical units $[\lambda]=[{\rm mass}]^{-2}$ and $[m]=[{\rm mass}]^{-1}$, 
such that the ratio $M_B\equiv \lambda^{\frac{1}{2}}/m^2$ 
has the dimension of a mass. We can then introduce 
the following dimensionless quantities:
\begin{subequations}
\begin{align}
\tilde{t}=t/M_B\quad\ ,&\quad \tilde{r}=r/M_B\,,\\
\tilde{P}=P M_B^{2}\quad\ ,\quad \tilde{\epsilon}=\epsilon &M_B^{2}\quad\ ,\quad \tilde{\omega}=\omega M_B\,.
\end{align}
\label{rescaling}
\end{subequations}
It is also convenient to define the dimensionless 
frequency $\tilde{\Omega}=\Omega/m\in(0,1)$, where for a BS
$\Omega$ is always smaller than $m$, and the limit $\tilde{\Omega}\to1$ 
corresponds to the non-relativistic (weak self-gravity) regime.

We can now scale the remaining dimensionless quantities 
by the ratio $\lambda^{1/2}/m$, 
in order to factor out the coupling constant 
from our equations:
\begin{equation}
\tilde{s}=\frac{m}{\lambda^\frac{1}{2}}s\quad\ ,\quad
|\tilde{\phi}|^2=\frac{\lambda}{m^2}|\phi|^2\ .
\label{stilde}
\end{equation}

Physical quantities can be restored after having solved the numerical problem by multiplying the dimensionless ones by different powers 
of $M_B$ to match the correct mass dimensions.

Hereafter we use $\tilde{s}$ as an input for the 
numerical integration of the field equations and threat it as a continuous parameter.
This is a valid approximation for configurations with large values of $s$ since, by virtue of the first relation in Eq.~\eqref{stilde}, the 
magnitude of $\tilde{s}$ will be large compared to the spacing 
between two consecutive values, since $m/\lambda^{1/2}\ll1$ 
\cite{ryan_spinning_1997}.
Solutions with a given $\tilde{s}$ can also be regarded as configurations with small $s$. This, however, implies a constraint on the physical masses and spins of the dimensionful 
rescaled configurations, since the ratio $\lambda^{1/2}/m$ 
becomes necessarily a multiple of $1/\tilde{s}$. Moderate and fast spinning BSs 
with small $s$ cannot be obtained with our method, because they will have 
large $\tilde{s}$ and the first equation in~\eqref{stilde} cannot be 
satisfied without violating the assumption $\lambda^{1/2}/m\gg 1$.
Indeed BSs with small $s$ and large $\chi$ only exist outside the 
strong self-coupling limit. 
With the variable transformations in Eqns.~\eqref{rescaling}-\eqref{stilde}, the problem translates 
in solving the Einstein equations for the metric specified 
by the line element 
\begin{align}
d\tilde{s}^2=&-e^{\gamma+\rho}d\tilde{t}^2+e^{2\alpha}(d\tilde{r}^2+\tilde{r}^2d\theta^2)\nonumber\\
&+e^{\gamma-\rho}\tilde{r}^2\sin{\theta}^2(d\varphi-\tilde{\omega} d\tilde{t})^2\ ,
\label{line_elem}
\end{align}
where the dimensionless metric functions $\rho,\gamma$, and $\alpha$ 
are the same as in Eq.~\eqref{Papape}. The stress-energy tensor in dimensionless variables reads
\begin{subequations}
\begin{align}
&\tilde{T}_{\mu\nu}=(\tilde{\epsilon}+\tilde{P})\tilde{u}_\mu \tilde{u}_\nu+\tilde{P}\tilde{g}_{\mu\nu}\,,\\
&(\tilde{u}_{\tilde{t}},\tilde{u}_{\tilde{r}},\tilde{u}_\theta,\tilde{u}_\varphi)=\frac{(-\tilde{\Omega},0,0,\tilde{s})}{(|\tilde{\phi}|^2+1)^{\frac{1}{2}}}\,,\\ 
&\tilde{\epsilon}=|\tilde{\phi}|^2+\frac{3}{4}|\tilde{\phi}|^4\quad\ ,\quad \tilde{P}=\frac{1}{4}|\tilde{\phi}|^4\,,
\label{rescaledenergy}
\end{align}
\end{subequations}
and the scalar field constraint becomes
\begin{equation}
|\tilde{\phi}|^2={\rm max}\left[0,\frac{(\tilde{\Omega}-\tilde{s}\tilde{\omega})^2}{e^{\gamma+\rho}}-\frac{e^{\rho-\gamma}\tilde{s}^2}{\tilde{r}^2\sin{\theta}^2}-1\right]\ .
\label{rescaledfield}
\end{equation}

For the sake of clarity, unless specified differently in the text, hereafter we shall drop the tilde 
from rescaled variables, and we will assume that all 
quantities are dimensionless.

\section{A self-consistent method for equilibrium configurations}\label{sec:selfmethod}

Finding BS solutions of the field equations 
\eqref{fieldequations} requires to solve an elliptic boundary 
value problem. To this aim we adopt the self-consisted method 
presented in~\cite{ryan_spinning_1997}, as an application of 
Hachisu self-consistent field approach.
The essence of this method lies in turning 
Einstein equations into an integral form which allows for 
an iterative resolution scheme. The first step toward 
the solution is writing the field equations in order to isolate 
on one side all operators having known Green functions and 
on the other side terms which can be regarded as 
effective sources.  

The Einstein equations for $\rho,\gamma$ and $\omega$ 
can be written in the following form 
\cite{komatsu_rapidly_1989}:
\begin{subequations}
\begin{align}
&\triangle(\rho e^{\frac{\gamma}{2}})=S_\rho(r,\mu)\,,
\label{firsteom}\\
&\left(\triangle+\frac{1}{r}\frac{\partial}{\partial r}-\frac{1}{r^2}\mu \frac{\partial}{\partial \mu}\right)\gamma  e^{\frac{\gamma}{2}}=S_\gamma(r,\mu)\,,
\label{secondeom}\\
&\left(\triangle+\frac{2}{r}\frac{\partial}{\partial r}-\frac{2}{r^2}\mu \frac{\partial}{\partial \mu}\right)\omega  e^{\frac{(\gamma-2\rho)}{2}}=S_\omega(r,\mu)\,,
\label{thirdeom}
\end{align}
\end{subequations}
where $\mu=\cos{\theta}$, $\triangle$ is the Laplacian 
operator in spherical coordinates, and the sources appearing on 
the right hand side are known expressions of the metric functions and their derivatives. 

The differential equation \eqref{firsteom} can be put in 
an integral form with the use of the three-dimensional 
Laplacian Green function:
\begin{equation}
\rho=-\frac{e^{-\frac{\gamma}{2}}}{4 \pi} \int_0^\infty dr' \int_{-1}^1 d\mu' \int_0^{2\pi} \frac{d\varphi' r'^2}{|r-r'|} S_\rho (r',\mu')\ .
\end{equation}

Using the expansion of $1/|r-r'|$ in powers of $r'/r$ (resp., $r/r'$), valid for $r'<r$ (resp., $r<r'$), one obtains 
the following integro-differential equation:
\begin{align}
\rho(r,\mu)=-&e^{-\gamma/2}\sum_{n=0}^{\infty}\int_0^\infty dr' R^n_\rho(r,r') \nonumber \\
&\times \int_0^1 d\mu' P_{2n}(\mu)P_{2n}(\mu')S_\rho(r',\mu')\,,
\label{rho}
\end{align}
where $P_{2n}(\mu)$ are the Legendre polynomials and
\begin{equation}
R^n_\rho(r,r')\equiv \frac{(r')^{2n+2}}{r^{2n+1}}\Theta(r'-r)+\frac{r^{2n}}{(r')^{2n-1}}\Theta(r-r')\,.
\label{rfunction}
\end{equation}
Analogous expressions can be found for Eqns.~\eqref{secondeom}-\eqref{thirdeom} with the same form as Eq.~\eqref{rho}:
\begin{align*}
f_i\propto e^{\hat{v}(f)}\sum_{n=1}^{\infty}\int_0^\infty dr' &R^n_{f_i}(r,r')\\ &\times\int_0^1 d\mu' {M}^n_{f_i}(\mu,\mu')S_{f_i}(r',\mu')\,,
\end{align*}
where $f=(\rho,\gamma,\omega)$, $\hat{v}$ is a linear function of $f$, $R^n_\gamma(r,r'), R^n_\omega(r,r') $ have the structure of Eq.~\eqref{rfunction} and ${M}^n_f(\mu,\mu')$ is an angular function including Legendre and associate Legendre polynomials. The asymptotic flatness conditions $\rho \sim \mathcal{O}(1/r)$, $\gamma \sim \mathcal{O}(1/r^2)$, $\omega \sim \mathcal{O}(1/r^3)$ for $r\to \infty$, are automatically satisfied if the source terms fall off sufficiently fast at large distances. We refer the reader to Appendix~\ref{metriceqns} for the full expression of the source terms. 

Finally, the remaining metric function $\alpha$ can be 
determined by integrating the differential equation 
\begin{equation}
\alpha_{,\mu}(r,\mu)=S_\alpha(r,\mu)\ ,
\label{eqalpha}
\end{equation}
together with the condition that $\alpha=\frac{1}{2}(\gamma-\rho)$ at the pole, where $S_\alpha(r,\mu)$ is given by Eq.~\eqref{alphaeq}.

\section{Multipole moments}
Relativistic multipole moments characterize 
the structure of astrophysical compact objects, their gravitational field, including non-linear contributions~\cite{hansen_multipole_1974} and their GW emission~\cite{PoissonWill}.
The actual computation of the multipole moments greatly simplifies 
in a wide class of \emph{asymptotically Cartesian and mass centered} 
coordinates~\cite{thorne_multipole_1980}, which allows reading the multipole moments directly off the asymptotic behavior of the 
metric coefficients.
Rotating axial (and equatorial) symmetry BSs 
are characterized by two families of scalar multipoles, the mass $\{M_{2i}\}_{i=0,\ldots\infty}$ 
and the current $\{S_{2i-1}\}_{i=1,\ldots\infty}$ moments, 
which can be extracted from the asymptotic behavior of the 
metric functions as in~\cite{ryan_spinning_1997}:
\begin{subequations}
\begin{align}
&\rho=-\sum_{n=0}^\infty\left[2 \frac{M_{2n}}{r^{2n+1}}
+\mathcal{O}\left(\frac{1}{r^{2n+2}}\right)\right]P_{2n}(\mu)\,, \label{multipolerho}\\
&\omega=-\sum_{n=1}^\infty\left[\frac{2}{2n-1}\frac{ S_{2n-1}}{r^{2n+1}}+\mathcal{O}\left(\frac{1}{r^{2n+2}}\right)\right]\frac{P_{2n-1}^1(\mu)}{\sin{\theta}}\,,
\label{multipoleomega}
\end{align}
\label{multipole}
\end{subequations}
with the lowest multipoles $M_0,S_1\equiv J$, and $M_2$ corresponding to the mass, 
angular momentum, and quadrupole moment, respectively. Comparing Eqns.~\eqref{multipole} 
with the explicit form of the metric in  Eqns.~\eqref{eqrho}-\eqref{eqomega}
it is straightforward to identify the mass and 
current moments 
as integrals over the source terms $S_\rho$ and $S_\omega$:
\begin{subequations}
\begin{align}
&M_{2n}=\frac{1}{2}\int_0^r dr' (r')^{2n+2}\int_0^1 d\mu' P_{2n}(\mu')S_\rho(r',\mu')\,,
\label{mass_mom}\\
&S_{2n-1}=\frac{1}{4n} \int_0^r dr' (r')^{2n+2}\nonumber\\
& \quad \quad
\times \int_0^1 d\mu' \sin{\theta'}P_{2n-1}^1(\mu')S_\omega(r',\mu')\,.
\label{current_mom}
\end{align}
\label{moments}
\end{subequations}

However, as noticed in~\cite{pappas_revising_2012}, the specific 
choice of radial coordinate leading to the line element \eqref{line_elem} 
renders the identifications of the multipole moments with the coefficients 
$M_{2n}$ and $S_{2n}$ ambiguous. To correctly match the definition of multipole moments given by Geroch and Hansen~\cite{hansen_multipole_1974}, all terms in 
Eq.~\eqref{multipole} with $n\ge 2$ must be corrected by adding a 
mass-spin dependent shift, yielding for the lowest moments: 
\begin{subequations}
\begin{align}
M_2^\tn{GH}=&M_2-\frac{4}{3}\Bigl(\frac{1}{4}+\frac{\gamma_0}{M_0^2}\Bigr)M_0^3\ ,\\
S_3^\tn{GH}=&S_3-\frac{12}{5}\Bigl(\frac{1}{4}+\frac{\gamma_0}{M_0^2}\Bigr)S_1 M_0^2\ ,
\end{align}
\label{multipole_corr}
\end{subequations}
where $M^\tn{GH}_{2n}$ and $S^{\rm GH}_{2n-1}$ are
Geroch-Hansen  moments, and the 
coefficient $\gamma_0$ can be read-off from the asymptotic 
$1/r$ expansion 
\begin{equation}
e^{\gamma}\sim\sqrt{\frac{\pi}{2}}\Bigl[\Bigl(1+\frac{\gamma_0}{r^2}\Bigr)T_0^{1/2}+\frac{\gamma_2}{r^4}T_2^{1/2}+....\Bigr]\ ,
\end{equation} 
where $T_l^{1/2}(\mu)$ are the Gegenbauer polynomials.
We discuss the relevance of such corrections in Sec.~\ref{sec:results}, 
but we can anticipate that, for all the BS configurations that we have considered, 
the correction due to the shift in \eqref{multipole_corr} is below $2\%$. For this reason, hereafter we will not distinguish 
between $\{M^{\rm GH}_{2n},S^{\rm GH}_{2n-1}\}$ and $\{M_{2n},S_{2n-1}\}$, 
discussing numerical results for the latter only.

It is convenient to introduce the \emph{reduced} multipoles of order $n$:
\begin{align}
\kappa_{2n}\equiv&(-1)^{n}\frac{M_{2n}}{\chi^{2n}M_0^{2n+1}} = (-1)^{n}\frac{\bar M_{2n}}{\chi^{2n}}\ ,\\
\sigma_{2n-1}\equiv&(-1)^{n+1}\frac{S_{2n-1}}{\chi^{2n-1}M_0^{2n}}=(-1)^{n+1}\frac{\bar S_{2n-1}}{\chi^{2n-1}}\ ,
\end{align}
with the leading multipoles being the reduced quadrupole and spin-octupole moments 
\begin{equation}
\kappa_2=-\frac{M_2}{\chi^2M_0^3}\quad , \quad \sigma_3=-\frac{S_3}{\chi^3M_0^4}\ . 
\end{equation}

These quantities are regular in the small-$\chi$ limit, and depend on the mass $M\equiv M_0$ and the effective coupling $M_B=\lambda^\frac{1}{2}/m^2$ only through the dimensionless combination $M/M_B$.
For a Kerr BH, $\kappa^\tn{BH}_{2n}=\sigma^\tn{BH}_{2n-1}=1$. 
As a comparison, for neutron stars $\kappa^\tn{NS}_{2}\sim1\div10$ 
depending on the internal composition~\cite{Urbanec:2013fs,Yagi:2016bkt}. Furthermore, for a Kerr BH $\kappa^\tn{BH}_{2n}$ and $\sigma^\tn{BH}_{2n-1}$ are independent of the spin, while for neutron stars this is true only to ${\cal O}(\chi^2)$.

\section{Numerical scheme}\label{Sec:scheme}

We have solved the system of equations for the metric functions 
and the scalar field discussed in Sec.~\ref{sec:selfmethod} by using a self-consistent iterative scheme. The full solution depends 
on the radius and the polar angle, defined on a 
two-dimensional grid $(r,\theta)$ with a fixed size 
(see discussion in the next section). Numerical calculations have 
been coded in \texttt{C} according to the following iterative procedure:
\begin{enumerate}
    \item We start by selecting an initial guess for the metric 
    functions  $(\rho,\gamma,\omega,\alpha)_{(1)}$, the angular 
    frequency $\Omega$ and a specific value of $s$. 
    The initial guess for the metric and the angular frequency corresponds
    to a solution of the field equations for a non-spinning, 
    spherically symmetric BS as explained in Appendix~\ref{guess}, while $s$ is 
    initialized to a small but non-zero value, $s\sim0.01$.
    \item From such initial configuration, we compute the energy density, pressure and scalar field amplitude $(\epsilon,P,\phi)_{(1)}$ using 
    Eqns.~\eqref{rescaledenergy}-\eqref{rescaledfield}.
    \item We replace $(\rho,\gamma,\omega,\Omega,\epsilon,P,\phi)_{(1)}$ 
    into the source terms on the right hand side of Eqns.~\eqref{metriceq}, 
    and perform the numerical integration, obtaining 
    the values of the metric functions at the next step 
    $(\rho,\gamma,\omega)_{(2)}$. The metric component $\alpha_{(2)}$ is 
    obtained by direct integration of Eq.~\eqref{eqalpha}. 
    \item The energy density, pressure and scalar field amplitude 
    $(P,\epsilon,\phi)_{(2)}$ are then computed from the above quantities, 
    thus completing one full iteration of the procedure. The solution
    is then improved iteratively by repeating steps $1$-$4$ until the desired
    convergence is reached.
    \item We use weighted averages 
    of the metric functions to boost the convergence of the algorithm. 
    Let $f_{(k)}$ collectively represent the values of each of the four components 
    $(\rho,\gamma,\omega,\alpha)$ after the $k_\tn{th}$ iteration. 
    Using $f_{(k)}$ to evaluate the source terms in Eq.~\eqref{metriceq} and integrating, we obtain the new values $\tilde{f}_{(k+1)}$, which one 
    would naively use as the inputs for the next iteration. Instead, following
    \cite{komatsu_rapidly_1989}, we build the linear combinations
    \begin{equation}
    f_{(k+1)}=a\tilde{f}_{(k+1)}+(1-a)f_{(k)}\ ,
    \end{equation}
    where $a\in(0,1)$ is a weight factor. 
    The use of weighted averages avoids the solution to bounce 
    among successive iterations. Hereafter we fix 
    $a=1/3$, which we found to provide the best compromise 
    between the speed and the accuracy of the convergence.
    \item   
    As a convergence criterion, we ask that
    the maximum relative difference between the values of all metric 
    functions at two successive iterations, evaluated on the two-dimensional 
    grid $(r,\theta)$, is smaller than a threshold $\delta$:
    \begin{equation}
        \qquad \Delta f=\max\limits_{(r,\theta)}
        \vert f_{(i+1)}/f_{(i)}-1\vert < \delta\,, \quad f=(\rho,\gamma,\alpha,\omega)\ .\label{error_iteration}
    \end{equation}
    For example, the algorithm needs about $150$ iterations for the solution to
    converge with a maximum relative error $\sim10^{-5}$.
\end{enumerate}
At each iteration, we adjust the input values of $\Omega$ and $s$
in such a way that the total mass and angular momentum, 
as determined by Eqns.~\eqref{moments}, are kept fixed to 
their predetermined desired values.
This adjustment is implemented through a two-dimensional 
Newton-Raphson method by solving the equations $M(\Omega_k,s_k)=M_\tn{fin}$ 
and $J(\Omega_k,s_k)=J_\tn{fin}$ for $(\Omega_k,s_k)$ at each $k$-th iteraction.
This procedure allows us to choose, at the beginning of 
the numerical simulation, the mass and spin of the BS solution\footnote{Keeping $M$ constant between 
different iterations is also necessary to guarantee 
convergence. Indeed, we found that leaving the value 
of $\Omega$ unchanged leads to a breakdown of the convergence after few iterations.}.

The metric functions are integrated on a 
two dimensional discrete grid for the coordinates $r$ and 
$\mu=\cos\theta$. We divide the angular domain into $n_\mu$ 
equally-spaced steps within $[0,1]$. We compactify the radial direction 
thorugh the change of coordinates
\begin{equation}
r\equiv r(q)=\frac{q}{1-q}\,,
\end{equation}
such that the radial domain $r\in[0,\infty)$ is mapped into
the finite domain $q\in[0,1)$.
We perform the integration between 
$q(r_0)=r_{0}/(r_{0}+1)$ and 
$q(r_\tn{\rm max})=r_\tn{\rm max}/(r_\tn{\rm max}+1)$, 
where $r_{0}=10^{-6}$ and $r_\tn{\rm max}=10$, which is typically two orders of magnitude larger than the radius of the BSs considered. 
We have verified that our results are stable for changes of 
both $r_0$ and $r_\tn{\rm max}$.
The number of grid points in the radial direction was fixed to $n_q=600$,  
while in the angular domain we choose different setups depending on the 
spin. 
Small values of $\chi$ render the metric profile stiffer, 
and require a more refined lattice with larger values 
of $n_\mu$. 

Derivatives in Eqns.~\eqref{sources} are numerically 
evaluated through a five-point central approximation, except
near the inner (outer) boundary were we used forward (backward)
derivatives. Integrals in Eqns.~\eqref{metriceq} 
are performed using the Simpson and the trapezoidal 
rule for the angular and the radial domain, respectively. 
We checked that higher order methods for both 
derivatives and integrals do not lead to significant 
changes in our results. 

Integration near the pole for Eqn.~\eqref{eqrho} and \eqref{eqomega} 
is simplified by resorting to the angular identities
\begin{align}
\lim_{\theta \to 0} \frac{\sin{[(2n-1)\theta]}}{(2n-1)\sin{\theta}}&=1\ ,\\    
\lim_{\theta \to 0} \frac{P^1_{2n-1}(\cos{\theta})}{2n(2n-1)\sin{\theta}}&=-\frac{(2n)!}{(n!)^22^{2n}(2n-1)} \ .
\end{align}

Finally, we fix the values of the components in the sum 
of Eq.~\eqref{metriceq} to $n=10$. 

\section{Results}\label{sec:results}

\subsection{Quadrupole and octupole moments of rotating massive BSs}
We have studied the multipolar structure of arbitrarily rotating massive BS in the large coupling limit
for different configurations specified by the spin parameter 
$\chi$ and by the object mass in units of $M_B$. To simplify 
the comparison with previous results in the literature, 
we include in our sample the range of masses considered 
in~\cite{ryan_spinning_1997} (we refer the reader to 
Appendix~\ref{app:comparions} for further comparisons).

We have carefully investigated how the obtained solutions 
are sensitive to the spacing of the numerical grid. We found that self-gravitating configurations are numerically stable under changes 
of the radial resolution (i.e. changes 
in $n_q$), while for 
small spins, typically $\chi\lesssim 0.1$, the integration 
becomes more sensitive to the angular resolution (i.e. to changes 
in $n_\mu$).
At large spins $\chi\sim\mathcal{O}(1)$, the radius, frequency, and
multipole moments are well determined and stable by choosing $n_q \sim n_\mu\sim \mathcal{O}(10^2)$ and 
setting $n=10$ in Eq.~\eqref{mass_mom}. 
Increasing the lattice density, as well as the cutoff 
value for $n$, typically yields changes of a few percent 
on the stellar structure. On the other hand, for slowly 
rotating BSs the calculation of multipole moments 
requires much larger values of $n_\mu\sim \mathcal{O}(10^4)$, to converge to a stable solution. For this reason, to extract the quadrupole and octupole moments,  we set the spacing of numerical grid to $n_q \times n_\mu = 600 \times 20000$ for $\chi\leq 0.1$ and $n_q \times n_\mu = 600 \times 1000$ for $\chi\geq 0.1$.

\begin{figure}[htp]
    \centering
    \includegraphics[width=0.49\textwidth]{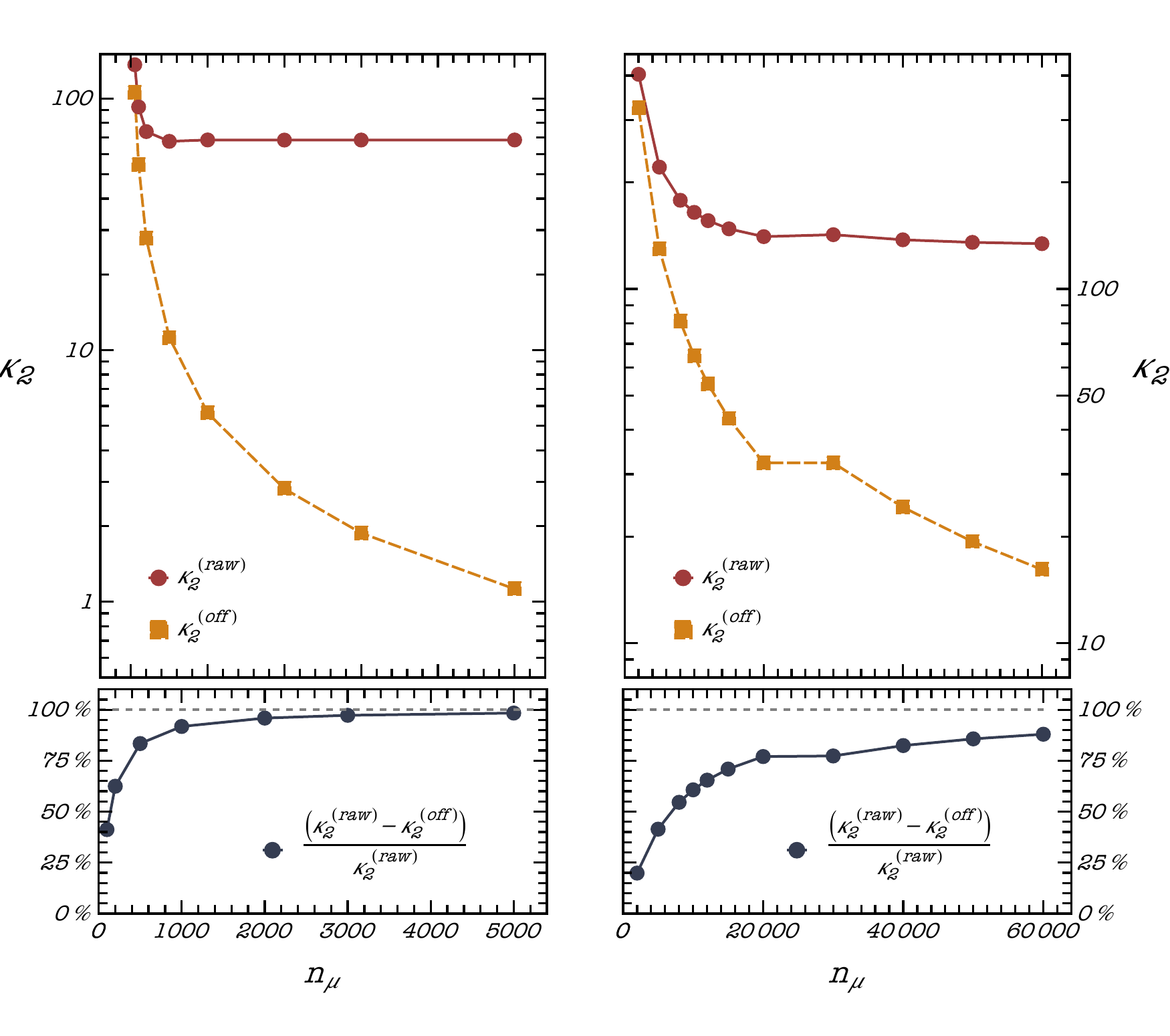}
    \caption{(Top panels) Data points identify raw values of 
    the reduced quadrupole moment, as well as the numerical 
    offset, as a function of the grid angular resolution. 
    (Bottom panels) Reduced quadrupole moment normalized by its 
    raw value as function of the angular resolution. Left and 
    right panels refer to BSs with spin $\chi=0.1$ and $\chi=0.0075$, respectively, both with $M/M_B=0.04$.}
    \label{fig:saturation}
\end{figure}

Moreover, as already discussed in~\cite{ryan_spinning_1997}, 
we find that for $\chi=0$ the quadrupole moment does 
not vanish, leading to a (small) numerical offset $M^\tn{(off)}_2=M_2(\chi=0)$. This is a numerical artifact, as we know that $M_2\sim \chi^2+{\cal O}(\chi^4)$. 
Therefore, we manually subtracted the offset from the raw values, i.e., we define the physical quadrupole moments as $M_2=M^\tn{(raw)}_2-M^\tn{(off)}_2$. 
The offset is negligible at large spin but it can spoil the $M_2\sim \chi^2$ scaling at small spins.
The top panels of Fig.~\ref{fig:saturation} show the reduced 
quadrupole moment $\kappa_2^\tn{(raw)}$ as obtained from the raw value of $M^\tn{(raw)}_2$, along with the corresponding value of the offset $\kappa^\tn{(off)}$, as a function of $n_\mu$ for a spinning BS with $\chi=0.1$ (left column) and $\chi=0.0075$ (right column). The mass of both configurations is fixed to $M/M_B=0.04$. 

While, for small $n_\mu$, $\kappa_2^\tn{(raw)}$ and $\kappa_2^\tn{(off)}$ have comparable magnitude, by increasing the value of $n_\mu$ the offset decreases monotonically and the effect of subtracting it becomes progressively less important. 
This is reflected in the bottom panels of Fig.~\ref{fig:saturation}, where we show that $\kappa_2/\kappa_2^\tn{(raw)}\to1$ as the angular resolution increases. The convergence is faster for higher spins (left panel). For a low value of the spin (right panel), the offset $\kappa_2^\tn{(off)}$ contributes $\approx25\%$ when $n_\mu=20000$, while the contribution reduces to $\approx12\%$ when $n_\mu=60000$, the convergence being monotonic with $n_\mu$. This is coherent with the fact, anticipated before, that convergence of the solution requires $n_\mu\simeq 10^3$ and $n_\mu\simeq 2\times10^4$ for $\chi\gtrsim 0.1$ and $\chi< 0.1$, respectively. 
We stress that only after subtracting the offset does $M_2$ scale as $\chi^2$ at small spins.

We speculate that the convergence of $\kappa_2$ with $n_\mu$ can be traced back to the BS topology. Solutions with small spin resemble closely the nonrotating spherical configurations. But, however small be the spin, rotating BSs are toroidal and they have no continuum limit to the spherical topology of the nonrotating case (because, for a given mass and fixed coupling, $\chi$ can only assume discrete values).
Configurations with spins close to the minimum 
value show a steep decrease of the energy density near the rotation axis, which requires a large number of angular points to be fully resolved. 
\begin{figure}[htp]
    \centering
    \includegraphics[width=0.49\textwidth]{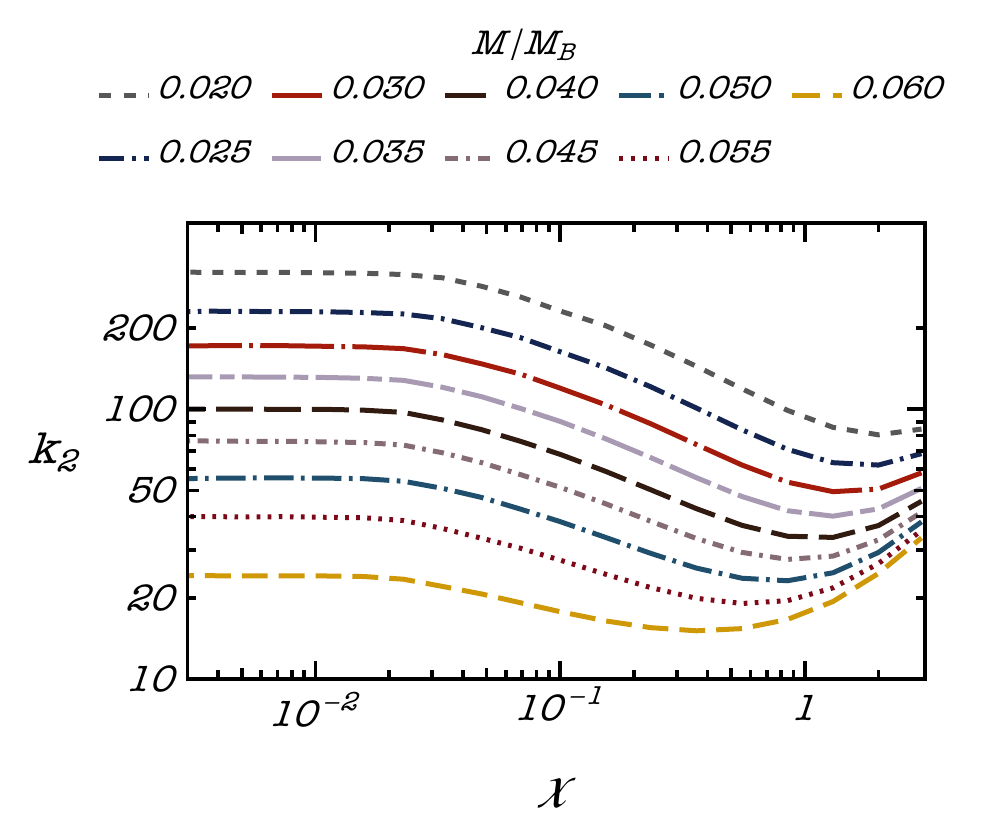}
    \caption{Reduced quadrupole moment $\kappa_2$ as a function 
    of the dimensionless spin $\chi$ for different values of the 
    BS mass in units of $M_B$. As a comparison, $\kappa_2^{\rm BH}=1$ for a Kerr BH with any spin.}
    \label{fig:Quadr_AllM}
\end{figure}

The reduced quadrupole moments $\kappa_2$ for different BS 
configurations, as a function of the dimensionless mass 
parameter $M/M_B$ and 
of the spin $\chi$, are shown in Fig.~\ref{fig:Quadr_AllM}.
For small values of $\chi$, $\kappa_2$ is nearly 
independent of the spin, i.e. $M_2\propto \chi^2$, 
with the proportionality constant depending only on 
the object mass. In the stable branch, configurations 
with larger masses are also more compact. Correspondingly, for fixed spin, $\kappa_2$ decreases as the mass increases. Interestingly, $\kappa_2$
is nonmonotonic with $\chi$, but it shows a gradual decrease between $\chi\sim0.03$ and $\chi\sim1$, after which it grows rapidly~\cite{ryan_spinning_1997}. Note that the spin can also exceed the Kerr bound, i.e. $\chi>1$.

The extraction of the reduced octupole moment $\sigma_3$ is more challenging due to the fact that, besides constant offsets, spurious numerical terms introduce additional nonphysical corrections at linear and quadratic order in the spin, spoiling the $\sigma_3\sim \chi^3+{\cal O}(\chi^5)$ dependence. Also in this case the offset is negligible for highly-spinning configurations.

In order to isolate the physical contribution, 
we fit the behavior of the octupole moment 
with a cubic polynomial $S^\tn{(raw)}_3(\chi)=a_0+a_1\chi+a_2\chi^2+a_3\chi^3$ 
for different small values of the spin parameter. For all 
BS configurations considered, we find non-zero 
values for the three coefficients $a_{0,1,2}$, with 
$a_0\sim a_2\ll a_1$.
After subtracting the constant, linear, and quadratic terms 
from the raw octupole moments, we recover the correct dependence $S_3\sim \chi^3$. Furthermore, to reduce the numerical noise, which can potentially affect the precision of the fit, we averaged over the last $50$ iterations of the algorithm, where $\Delta f$ in Eq.~\eqref{error_iteration} oscillates about its minimum value. As for $M_2$, we find that the spurious coefficients $a_{0,1,2}$ decrease for higher angular resolution (i.e., for larger values of $n_\mu$).

\begin{figure}[htp]
    \centering
    \includegraphics[width=0.49\textwidth]{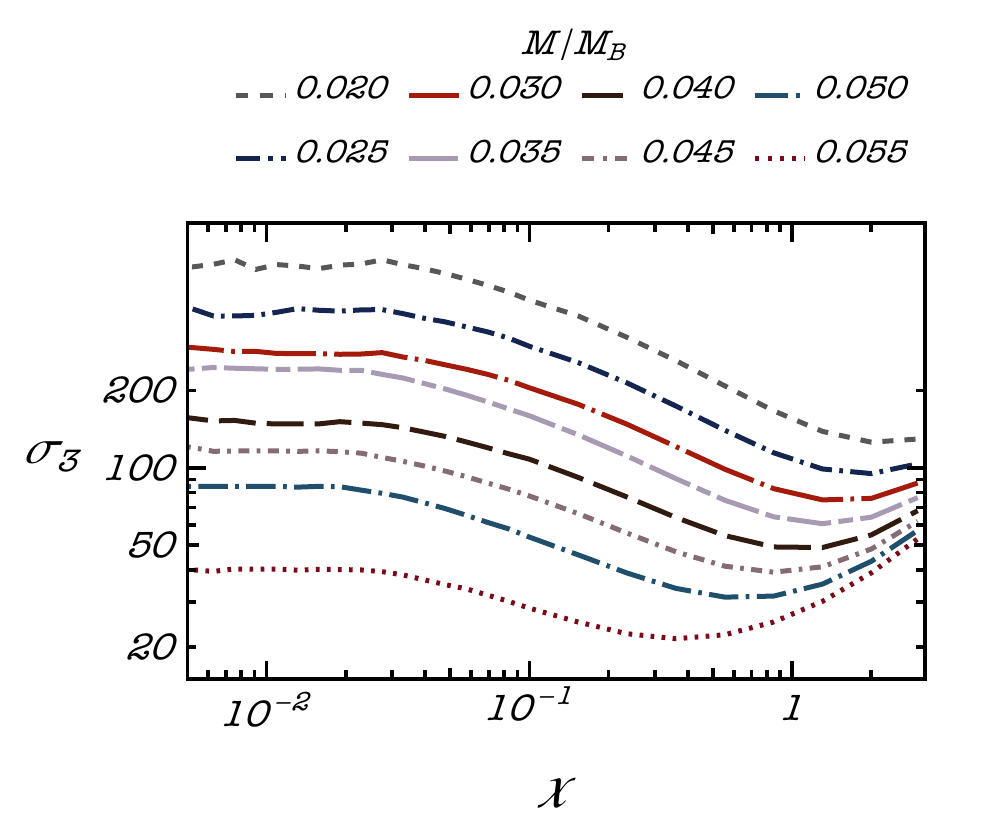}
    \caption{Reduced spin-octupole moment $\sigma_3$ as a function of the dimensionless spin $\chi$ for the same BS masses of Fig.~\ref{fig:Quadr_AllM}. As a comparison, $\sigma_3^{\rm BH}=1$ for a Kerr BH with any spin.}
    \label{fig:octupole}
\end{figure}

However, the extraction of $\sigma_3$ is problematic for masses close to $M/M_B\simeq0.06$, i.e., to the maximum mass of non rotating BSs. As already observed in~\cite{ryan_spinning_1997}, the octupole moment is small for such masses and its accurate determination is prevented by numerical uncertainties. For this reason, we do not report the corresponding data. For the other configurations analyzed in Fig.~\ref{fig:Quadr_AllM}, the reduced spin-octupole moment $\sigma_3=-S_3/\chi^3M^4$,
obtained though the procedure described above, is shown in Fig.~\ref{fig:octupole}.
As for the quadrupole moments, the curves are constant at 
small spins and exhibit a transition to a region with negative 
slope in correspondence roughly of the same values of $\chi$. 

The data for the quadrupole and octupole as a function of the spin $\chi$ and mass $M/M_B$ are publicly available online~\cite{webpage}.

The values for $\kappa_2$ and $\sigma_3$ plotted above ignore the corrections in the definition of the Geroch-Hansen multipole moments, Eq.~\eqref{multipole_corr}. We show that, indeed, these corrections introduce a shift at the (sub-)percent level and therefore they can be ignored at the current numerical precision.
As a representative example, we focus on a specific BS configuration with mass 
$M=0.06M_B$ and different values of the spin. The corrections 
to the reduced quadrupole are shown in the last column of 
Table~\ref{tab:fiducial} for such models.
\begin{table}[htp]
  \begin{tabularx}{0.9\linewidth}{
  >{\hsize=.8\hsize\linewidth=\hsize}X
>{\hsize=.8\hsize\linewidth=\hsize}X
>{\hsize=.8\hsize\linewidth=\hsize}X
>{\hsize=1.6\hsize\linewidth=\hsize}X}
    \hline\hline
     \rule{0pt}{2.5ex} 
        $\chi$ &$\kappa_2$ &$\kappa_2^{\rm new}$ &${\rm correction}[\%]$\\
         \hline
         $0.1$ &$22.4$ &$22.1$ &$\quad -1.4\%$ \\
         $0.2$ &$15.7$ &$15.6$ &$ \quad -0.5\%$\\
         $0.5$ &$15.2$ &$15.3$ &$\lesssim+0.1\%$ \\
         $0.8$ &$16.4$ &$16.4$ &$\lesssim+0.1\%$\\
         $1.0$ &$17.4$ &$17.5$ &$\lesssim+0.1\%$ \\
         $1.3$ &$19.3$ &$19.4$ &$\lesssim+0.1\%$ \\
         $2.0$ &$24.6$ &$24.6$ &$\lesssim+0.05\%$\\
     \hline\hline
  \end{tabularx}
    \caption{Corrections to the reduced quadrupole moment 
    derived in~\cite{pappas_revising_2012}, for different 
    value of the spin $\chi$ and $M=0.06M_B$.}
    \label{tab:fiducial}
\end{table}
We find that corrections to $\kappa_2$ are in general small, 
never exceeding a relative difference $\sim 2\%$, for the 
whole range of spins considered. The correction is larger for 
more compact configurations, therefore, given that 
$M=0.06M_B$ corresponds 
to the maximum value of the compactness for non-spinning BSs, 
changes in $\kappa_2$ are even smaller for 
lower values of the mass, as those analysed in Fig \ref{fig:Quadr_AllM}. 
This picture holds as well for $\sigma_3$, for which we find 
corrections smaller than $0.6\%$ for configurations near the maximum considered mass. 

Finally, we have also checked that the first moments, namely the mass and spin as computed from Eqs.~\eqref{multipole} agree with those obtained from the Komar integrals \cite{poisson_relativists_2004}:
\begin{align*}
M&=-8\pi \int_0^\infty dr~\int_0^1 d\cos{\theta} r^2e^{2\alpha+\gamma}\Bigl(T^t_t t^t-\frac{1}{2}Tt^t\Bigr) \,,\\
J&=4\pi \int_0^\infty dr~\int_0^1 d\cos{\theta} r^2e^{2\alpha+\gamma}T^t_\varphi\,.
\end{align*}

\subsection{Maximum mass, compactness, ergoregions}

Together with the multipolar structure, our framework 
allows describing various features of rotating BSs, 
such as the dependence of the maximum mass and of the 
compactness on the spin and frequency, as well as the presence of ergoregions.

\begin{figure}[htp]
    \centering
    \includegraphics[width=0.49\textwidth]{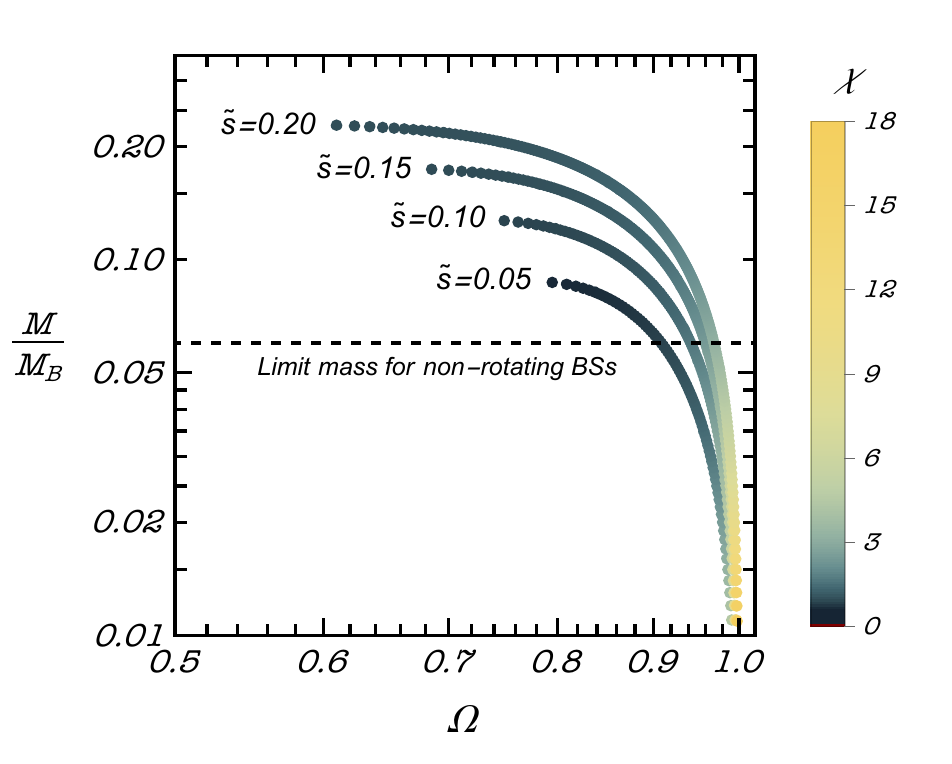}
    \caption{
    Boson star mass as a function of the 
    frequency $\Omega$ for four different values of 
    $\tilde{s}$. The color range of each configurations 
    is mapped to the value of the $\chi$, 
    with, darker (lighter) tones corresponding to smaller 
    (larger) spins. }
    \label{fig:max_mass}
\end{figure}

Due to centrifugal forces which work against the gravitational 
collapse, rotating BSs can support larger masses, 
compared to their spherically symmetric counterparts,   
as also shown by the mass-frequency curves in 
Fig.~\ref{fig:max_mass} for four representative 
families of solutions with different value of $\tilde{s}$
\footnote{As explained in Sec.~\ref{Sec:scheme} 
our code uses $\chi$ as input parameter. However, for 
the maximum mass analysis, we have changed the workflow 
in order to have $\tilde{s}$, together with the BS mass, 
as input. We also remark that, as discussed in Sec.~\ref{Sec:rescaling}, 
the rescaled winding number $\tilde{s}$ does not need to 
be an integer and it depends on the coupling constants $\lambda$ and $m$ 
as in Eq.~\eqref{stilde}.}. 

For a given value of $\tilde{s}$ the mass of each sequence of 
solutions grows as $\Omega$ decreases, until the 
maximum mass is reached, which is identified in our code 
by a failure of the algorithm to converge.
Previous studies, which focused on massive BSs with 
non-rescaled winding number $s=1$, showed that such sequences are continuously connected 
for smaller frequencies to linearly unstable branches, 
in which $dM/d\Omega<0$~\cite{herdeiro_kerr_2015}.

Figure \ref{fig:max_mass} also shows the values of $\chi$ for 
the different configurations. Families of solutions 
with large $\tilde{s}$ have high spins as long as their frequency 
remains large. In particular, note that also configurations with $\chi\gg1$ are allowed.
As $\Omega$ decreases along the curve, 
the mass and compactness increase and $\chi$ rapidly falls, 
approaching a value $\chi(M_{\rm max})<1$. Moreover, 
$\chi(M_{\rm max})\simeq 1$ for all stars with 
$\tilde{s}$ large enough that significative rotation rate and compactnesses are approached along the curve. 

\begin{figure}[htp]
    \centering
    \includegraphics[width=0.49\textwidth]{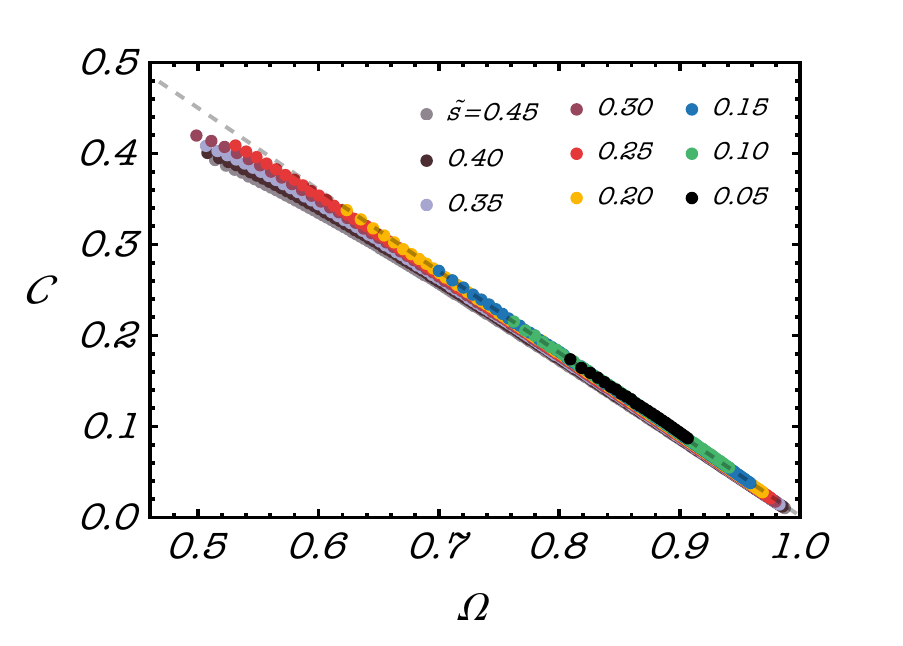}
    \caption{Colored dots identify the BS compactness as a function 
    of the frequency for stars with $\tilde{s}$ in the range $(0.05,0.45)$. 
    The mass for each family, varies between $M=0.06M_B$ and 
    the maximum mass allowed by $\tilde{s}$. The dashed black 
    line corresponds to a linear fit of the data for 
    $\tilde{s}\leq 0.2$, i.e. ${\cal C}=0.9(1-\Omega)$.}
    \label{fig:comp}
\end{figure}

Beside the maximum mass, we have also analysed 
the dependence on $\Omega$ of the BS compactness
${\cal C}=M/R$, with
\begin{equation*}
R=R_0e^{\rho(R_0,\pi/2)-\gamma(R_0,\pi/2)}\ ,
\end{equation*}
being the perimetral radius and $R_0$ the stellar 
radius, i.e. the value of the $r-$coordinate for which 
the scalar field vanishes, marking the division between 
the tail and the non-tail region (see Sec.~\ref{sec:tails}).
Figure \ref{fig:comp} shows ${\cal C}$ as a function of the 
frequency, for the same stellar configurations 
considered before, plus other five with larger $\tilde{s}$.

Interestingly, for $\tilde{s} \lesssim 0.2$, the compactness 
depends linearly on the frequency and the relation is 
independent of the value of $\tilde{s}$ (or $\chi$). The latter only affects the minimum value of the frequency which can be reached by each family 
and the mass profile. This linear relation holds also for mini BSs in the stable branch, as can be appreciated examining the data in~\cite{delgado_rotating_2020}.

All families with $\tilde{s}\gtrsim 0.25$ reach a 
maximum value ${\cal C}_{\rm max}\simeq 0.4$, smaller 
than the Buchdhal limit\footnote{Note that 
although massive BSs in the strong coupling limit are 
described by a perfect fluid stress energy tensor, the 
Buchdahl limit~\cite{Buchdahl:1959zz} does not apply due to rotation~\cite{Cardoso:2019rvt}.}, around $\Omega_{\rm min} \simeq 0.5$.

\begin{figure}[htp]
    \centering
    \includegraphics[width=0.49\textwidth]{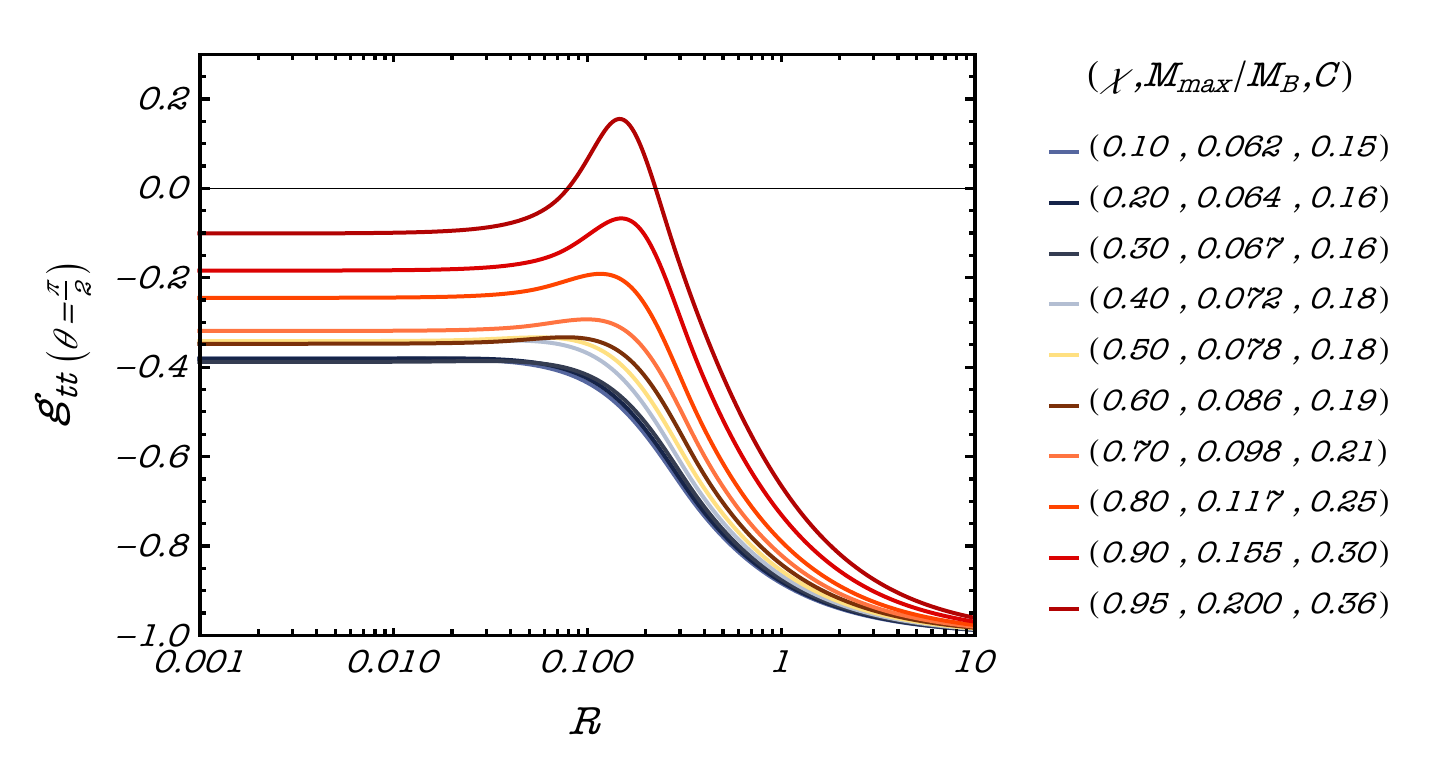}
    \caption{Metric component $g_{tt}$ as function of 
    the radius on the equatiorial plane, for 
    configurations corresponding to the maximum mass 
    at a given spin $\chi$. For $(\chi,M)=(0.95,0.2M_B)$ 
    the $g_{tt}$ changes sign twice, revealing the 
    presence of an ergoregion with toroidal topology.}
    \label{fig:ergoreg}
\end{figure}

Due to their large compactness, it is reasonable to expect that 
massive and fast spinning configurations develop ergoregions. 
Figure~\ref{fig:ergoreg} shows indeed a sequence of BSs at the 
maximum mass allowed for a given value of $\chi$, 
which feature an ergoregion for sufficiently high spin. Notice that the first appearance of an ergoregion is for a configuration with $\chi \gtrsim 0.9$, ${\cal C}\gtrsim 0.30$. 
Such a BS has $\tilde{s}\sim 0.1$, which translates, 
in the $\lambda^{1/2}/m\gg1$ limit, to a winding number 
$s\gg 1$. The ergoregion shown in Fig.~\ref{fig:ergoreg} 
arise for solutions in the stable branch\footnote{This is different 
from  the case of mini BSs, 
which exhibit ergoregions only for configurations in the unstable branch~\cite{delgado_rotating_2020}.}.
However, although stable against radial perturbations, BSs with an ergoregion are unstable over longer timescales against nonspherical modes due to the so-called  ergoregion instability~\cite{1978CMaPh..63..243F,Cardoso:2007az}. An interesting followup of our work could be to quantify the instability time scale for our configurations.

\section{Conclusions and discussion}
In this work we constructed fully relativistic solutions of rotating BSs with quartic self-interactions within a perfect fluid approximation scheme, valid in the large self-coupling regime. The Einstein equations have been solved with a numerical \texttt{C} code implementing the iterative method described in~\cite{ryan_spinning_1997}, which allows us to find configurations covering a wide portion of the parameter space, including those which are more relevant for phenomenology. Indeed, since the coupling constants are completely factored out from the numerical solution, each configuration corresponds to a family of BSs, sharing the same compactness and dimensionless spin but differing in the mass, the latter scaling linearly with the combination of the self-coupling and the boson's mass defined in Sec.~\ref{Sec:rescaling}. 

We characterized the multipolar structure of these BSs up to the spin-octupole contribution, considering different sequences of compact configurations with constant mass, spanning a two-dimensional region in the mass-spin parameter space, including the slowly rotating regime. The values of the quadrupole and spin-octupole moments have been computed significantly more accurately than in previous work.

Our results, summarized in Fig.~\ref{fig:Quadr_AllM} and Fig.~\ref{fig:octupole}, confirm that the quadrupole moment is proportional to $\chi^2$ (as in the Kerr case) but only for slowly spinning BSs and that the constant reduced quadrupole moment in that regime has a minimum, causing the range of values of $\kappa_2$ to be not continuously connected to the BH value $\kappa_2^{\rm BH}=1$.  
We found such minimum, as well as the reduced multipoles, except for very large spin configurations, to be larger than what reported in previous work. We also confirmed that the spin-octupole is proportional to $\chi^3$ (as in the Kerr case) only for low spins. 

Moreover, we discussed the masses and compactness of these objects, analyzing solutions with fixed rescaled winding number $\tilde{s}$ and for given values of the coupling constants. We showed that the maximum BS mass increases considerably for high $\tilde{s}$ and, as it grows, the maximum mass configuration is reached for lower and lower frequencies. The corresponding compactnesses approaches $\mathcal{C}\sim 0.4$, while the  dimensionless spin parameter $\chi$ is close to unity. We found that some of these configurations have ergoregions in the branch connected to the Newtonian limit $\Omega \to 1$. 

\begin{figure}[htp]
    \centering
    \includegraphics[width=8cm]{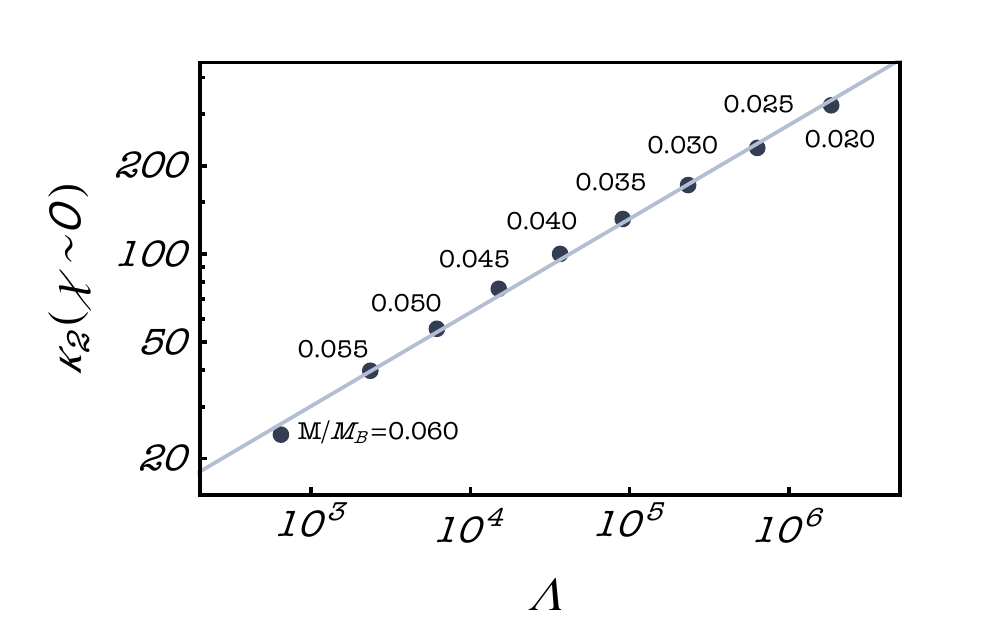}
    \caption{Reduced quadrupole moment $\kappa_2$ as a function of the 
    tidal deformability for different values of the BS mass, 
    in the low spin region, $\chi\lesssim 0.02$.}
    \label{fig:loveq}
\end{figure}

Among various theoretical and observational applications of our results, multipole moments have interesting phenomenological consequences related to the so-called universal relations~\cite{Yagi:2016bkt}. Indeed, it is known  that approximated analytical relations exist between certain observables of a neutron star, such as the spin-induced 
quadrupole moment, the tidal deformability $\Lambda$, and the moment of inertia, which are roughly insensitive of the 
underlying equation of state~\cite{yagi_i-love-q_2013}. 
The same functional form of these relations holds also for the reduced quadrupole moment of slowly spinning massive 
BSs and its corresponding $\Lambda$, as
we have recently shown~\cite{pacilio_gravitational-wave_2020}. 
The numerical calculations discussed in this paper allows 
to strengthen our previous result, obtained with limited 
data and accuracy. Using the fits for $\Lambda$ provided in Ref.~\cite{Sennett:2017etc}, the $\kappa_2-\Lambda$ relation is shown Fig.~\ref{fig:loveq}, with the straight line identifying the semi-analytical fit 

\begin{equation}
\log \kappa_2\simeq 1.2+0.32 \log\Lambda \ .
\end{equation}
We measured the distance of the data from the fit as the root mean square relative error $\sigma_E=0.01$, where $\sigma_E^2=\frac{1}{N}\sum_{n=1}^N (r_E^i)^2$ and $r_E^i=\kappa_2^i-(1.2+0.32\Lambda^i)$ are the residuals.

\begin{figure}[htp]
    \centering
    \includegraphics[width=8cm]{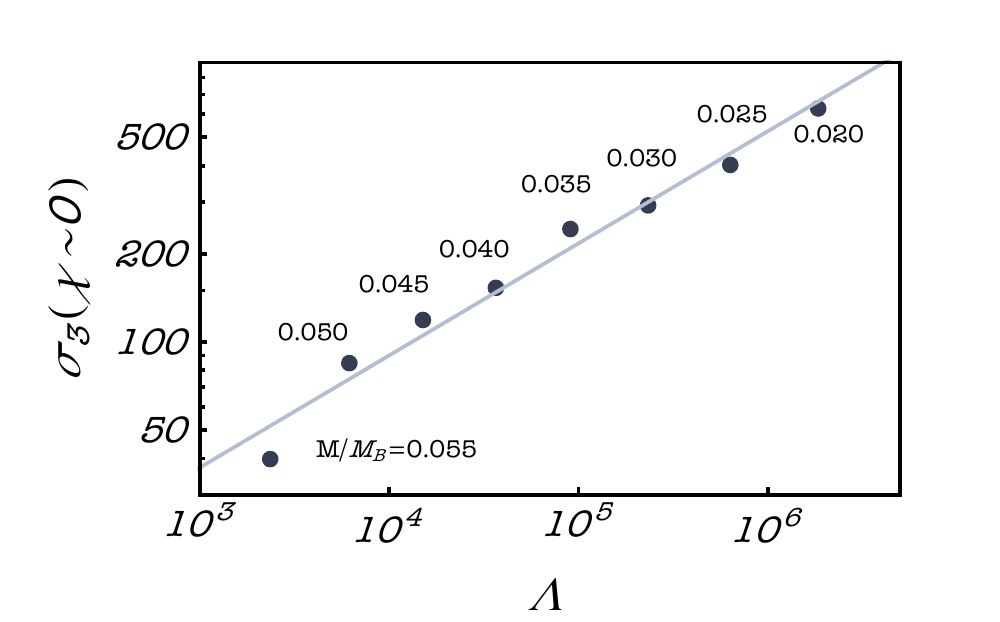}
    \caption{Reduced spin-octupole moment $\sigma_3$ as a function of the 
    tidal deformability for different values of the BS mass, 
    in the low spin region, $\chi\lesssim 0.02$. As for the quadrupole in Fig.~\ref{fig:loveq}, 
    we show the best-fit line, corresponding to $\sigma_3 \simeq 0.98 + 0.38 \log \Lambda$. The distance from the fit is measured by $\sigma_E=0.03$.}
    \label{fig:loveoct}
\end{figure}

We also used the data in Fig.~\ref{fig:octupole} to explore
the $\sigma_3 - \Lambda$ relation. To reduce the numerical noise of $\sigma_3$ at low spins, 
we average $\sigma_3(\chi \sim 0)$ over $10$ points for $\chi \lesssim 0.02$, for each value of the mass.
The data are shown in Fig.~\ref{fig:loveoct} and suggest that a simple linear relation exists as well between $\log\kappa_2$ and $\log\Lambda$.
The relations discussed above might have many applications and 
are especially useful to break degeneracies among 
parameters that characterise gravitational waveforms~\cite{pacilio_gravitational-wave_2020}. On the theoretical 
side, proving that such $\kappa_2-\Lambda$ and $\sigma_3-\Lambda$ 
relations also exist 
for other scalar field interactions is an interesting 
and challenging task, which will shed light on the 
origin of the universality, and will be investigated in a followup publication.
The results presented in this work are valid as long as the self-coupling is large, in which case the anisotropies of the star are negligible.  Nonetheless the approach is not limited to BSs with $s\gg1$, even if, for small $s$, only slowly rotating configurations can fully satisfy the requirement $\lambda/m^2\gg1$ coherently with the first of Eqs.~\eqref{stilde}.
The extension of these results to the generic coupling regime and to different BSs potentials will be explored elsewhere.
Likewise, it would be interesting to extend our analysis to compute the multipolar structure of Proca stars~\cite{Brito:2015pxa} or of Kerr BHs with bosonic hair~\cite{Herdeiro:2014goa,Herdeiro:2015waa,Herdeiro:2015gia,Herdeiro:2016tmi}.

\begin{acknowledgments}
We are indebted to Carlos Herdeiro for discussions 
and comments on the manuscript. Numerical calculations have been made possible through a CINECA-INFN agreement, providing access to resources on MARCONI at CINECA. We acknowledge
financial support provided under the European Union's H2020 
ERC, Starting Grant agreement no.~DarkGRA--757480. We 
also acknowledge support under the MIUR PRIN and FARE 
programmes (GW-NEXT, CUP:~B84I20000100001), and 
from the Amaldi Research Center funded by the MIUR 
program ``Dipartimento di Eccellenza'' 
(CUP: B81I18001170001).
\end{acknowledgments}

\appendix

\section{Field equations for arbitrarily spinning BSs in the large coupling limit}\label{metriceqns}
In this Appendix we provide the full integral form of 
the equations for the metric functions $\rho, \gamma$ and $\omega$, derived in~\cite{komatsu_rapidly_1989}: 
\begin{widetext}
\begin{subequations}
\begin{eqnarray}
\rho(r,\mu)=-e^{-\gamma/2}\sum_{n=0}^{\infty}P_{2n}(\mu)\Bigg[ \frac{1}{r^{2n+1}}\int_0^r dr' (r')^{2n+2}\int_0^1 d\mu' P_{2n}(\mu')S_\rho(r',\mu')\nonumber \\
+r^{2n} \int_r^\infty dr' \frac{1}{(r')^{2n-1}} \int_0^1 d\mu' P_{2n}(\mu')S_\rho(r',\mu') \Bigg]\ ,
\label{eqrho}
\end{eqnarray}
\begin{eqnarray}
\gamma(r,\mu)=-\frac{2}{\pi} e^{-\gamma/2} \sum_{n=1}^\infty \frac{\sin{[(2n-1)\theta]}}{(2n-1)\sin{\theta}} \Bigg[ \frac{1}{r^{2n}} \int_0^r dr' (r')^{2n+1} \int_0^1 d\mu' \sin{[(2n-1)\theta']} S_\gamma (r',\mu')\nonumber\\
+r^{2n-2} \int_r^\infty dr' \frac{1}{(r')^{2n-3}} \int_0^1 d\mu' \sin{[(2n-1)\theta']}S_\gamma(r',\mu')\Bigg]\ ,
\label{eqgamma}
\end{eqnarray}
\begin{eqnarray}
\omega(r,\mu)=-e^{\rho-\gamma/2} \sum_{n=1}^\infty \frac{P_{2n-1}^1(\mu)}{2n(2n-1)\sin{\theta}}\Bigg[\frac{1}{r^{2n+1}} \int_0^r dr' (r')^{2n+2} \int_0^1 d\mu' \sin{\theta'}P_{2n-1}^1(\mu')S_\omega(r',\mu')\nonumber\\
+r^{2n-2} \int_r^\infty dr' \frac{1}{(r')^{2n-3}} \int_0^1 d\mu' \sin{\theta'} P_{2n-1}^1 (\mu') S_\omega(r'\mu')\Bigg]\ .
\label{eqomega}
\end{eqnarray}
\label{metriceq}
\end{subequations}
\end{widetext}
The functions $P_n(\mu)$ and $P_n^m(\mu)$ correspond to the Legendre and associate Legendre polynomials, respectively.
The sources, defined in 
Eqns.~\eqref{firsteom}-\eqref{thirdeom}, read
\begin{widetext}
\begin{subequations}
\begin{eqnarray}
S_\rho(r,\mu)=e^{\gamma/2}\Bigg\{ 8\pi e^{2\alpha} (\epsilon+P)\frac{1+v^2}{1-v^2}+r^2(1-\mu^2) e^{-2\rho} \Bigg(\omega^2_{,r}+\frac{1-\mu^2}{r^2}\omega^2_{,\mu} \Bigg)+\frac{1}{r}\gamma_{,r}-\frac{\mu}{r^2} \gamma_{,\mu}\nonumber\\
 +\frac{1}{2} \rho \Bigg[16\pi e^{2\alpha}P-\gamma_{,r}\Bigg(\frac{1}{2}\gamma_{,r}+\frac{1}{r}\Bigg)-\frac{1}{r^2} \gamma_{,\mu} \Bigg(\frac{1-\mu^2}{2} \gamma_{,\mu} -\mu\Bigg) \Bigg] \Bigg\}\ ,
\end{eqnarray}
\begin{eqnarray}
S_\gamma(r,\mu)=e^{\gamma/2}\Bigg[16\pi e^{2\alpha}P+\frac{\gamma}{2} \Bigg(16 \pi e^{2\alpha} P-\frac{1}{2} \gamma^2_{,r}-\frac{1-\mu^2}{2r^2} \gamma^2_{,\mu}\Bigg)\Bigg]\ ,
\end{eqnarray}
\begin{eqnarray}
S_\omega(r,\mu)=e^{\gamma/2-\rho}\Bigg\{ -16\pi e^{2\alpha+\rho}\frac{v(\epsilon+P)}{(1-v^2)r\sin{\theta}}+\omega \Bigg[-8\pi e^{2\alpha} \frac{(1+v^2)\epsilon +2v^2P}{1-v^2} -\frac{1}{r}(2\rho_{,r}+\frac{1}{2}\gamma_{,r}) \nonumber \\
+\frac{\mu}{r^2}(2\rho_{,\mu}+\frac{1}{2}\gamma_{,\mu})+\rho^2_{,r}-\frac{1}{4}\gamma^2_{,r}+\frac{1-\mu^2}{r^2}(\rho^2_{,\mu}-\frac{1}{4}\gamma^2_{,\mu})-r^2(1-\mu^2)e^{-2\rho}\Bigg(\omega^2_{,r}+\frac{1-\mu^2}{r^2}\omega^2_{,\mu}\Bigg)\Bigg]\Bigg\}\ .
\end{eqnarray}
\label{sources}
\end{subequations}
\end{widetext}
The parameter $v$ entering in the previous expressions can be 
identified as the proper velocity with respect to the 
zero angular momentum observer and is given by:
\begin{equation}
v=\frac{s}{\Omega-s\omega} \frac{e^{\rho}}{r\sin{\theta}}\ .
\end{equation}
Finally, the function $\alpha$ can be determined by solving
\begin{widetext}
\begin{eqnarray}
\alpha_{,\mu}=-\frac{1}{2}(\rho_{,\mu}+\gamma_{,\mu})+\{\frac{1}{2}[r^2(\gamma_{,rr}+\gamma^2_{,r})-(1-\mu^2)(\gamma_{,\mu\mu}+\gamma^2_{,\mu})][-\mu+(1-\mu^2)\gamma_{,\mu}] \nonumber \\
+r\gamma_{,r}[\frac{1}{2}\mu+\mu r \gamma_{,r}+\frac{1}{2}(1-\mu^2)\gamma_{,\mu}]+\frac{3}{2}\gamma_{,\mu}[-\mu^2+\mu(1-\mu^2)\gamma_{,\mu}]\nonumber \\
-r(1+r \gamma_{,r})(1-\mu^2)(\gamma_{,r\mu}+\gamma_{,r}\gamma_{,\mu})-\frac{1}{4}\mu r^2 (\rho_{,r}+\gamma_{,r})^2-\frac{1}{2}r (1+r\gamma_{,r})(1-\mu^2)(\rho_{,r}+\gamma_{,r})(\rho_{,\mu}+\gamma_{,\mu})\nonumber \\
+\frac{1}{4}\mu (1-\mu^2)(\rho_{,\mu}+\gamma_{,\mu})^2+\frac{1}{4}r^2 \mu (1-\mu^2)\gamma_{,\mu}[r^2(\rho_{,r}+\gamma_{,r})^2-(1-\mu^2)(\rho_{,\mu}+\gamma_{,\mu})^2] \nonumber \\
+(1-\mu^2) e^{-2\rho}(\frac{1}{4}r^4\mu \omega^2_{,r}+\frac{1}{2}r^3(1-\mu^2) \omega_{,r}\omega_{,\mu}-\frac{1}{4}r^2 \mu(1-\mu^2) \omega^2_{,\mu}+\frac{1}{2}r^4(1-\mu^2)\gamma_{,r}\omega_{,r}\omega_{,\mu}\nonumber \\
-\frac{1}{4} r^2(1-\mu^2)\gamma_{,\mu}[r^2 \omega^2_{,r}-(1-\mu^2)\omega^2_{,\mu}])\}/\{(1-\mu^2)(1+r\gamma_{,r})^2+[\mu-(1-\mu^2)\gamma_{,\mu}]^2\}\ .
\label{alphaeq}
\end{eqnarray}
\end{widetext}
with appropriate boundary conditions, which correspond to \eqref{eqalpha} with $S_\alpha(r,\mu)$ written explicitly on the right hand side. 

\section{Initial data for rotating configurations}\label{guess}
The self-consistent iterative scheme to build 
spinning BS solutions requires an initial guess for 
the metric functions $(\rho,\gamma,\omega,\alpha)$ 
and the frequency $\Omega$. For such initial 
data, we choose a solution describing a nonrotating BS 
with the same mass of the rotating configuration we 
want to obtain. \\
In the non-spinning limit, $\tilde{\omega} \to 0$, 
$e^{\gamma-\rho}=e^{2\alpha}$ and the metric reduces to
\begin{equation}
 d\tilde{s}^2=-e^{2(\rho+\alpha)}d\tilde{t}^2+e^{2\alpha}(d\tilde{r}^2+\tilde{r}^2d\theta^2+\tilde{r}^2\sin{\theta}^2d\phi^2)\ ,
 \label{quasiisometric}
\end{equation}
in which the metric functions $\rho$ and $\alpha$ are independent of the angular variable 
$\mu=\cos\theta$. However, for spherically symmetric solutions, it proves useful to use a metric ansatz expressed in Schwarzschild-like coordinates:
\begin{equation}
ds^2=-e^{v(r)}dt^2+e^{u(r)}dr^2+r^2d\theta^2+r^2\sin{\theta}^2d\phi^2\ .
\label{polarmetric}
 \end{equation}
A relation between the metric functions $(\rho,\gamma,\alpha)$ 
and $(u,v)$ can be found once we determine a coordinate 
transformation that maps metric \eqref{polarmetric} 
into Eq.~\eqref{quasiisometric}. Let's start observing 
that assuming $\tilde{t}=t$ we have $\rho(\tilde{r})+\alpha(\tilde{r})=v(r)/2$. Moreover, 
from the spatial components of the metric:

\begin{equation}
 \frac{d\tilde{r}}{r}=\frac{e^{\frac{u(r)}{2}}}{r}dr\ ,
\end{equation}
which integrated, gives the desired map
\begin{equation}
\tilde{r}(r)\propto\cdot \tn{exp}\Bigl[\int_{r_0}^r\frac{e^{\frac{u(r')}{2}}}{r'}dr'\Bigr]\ .
\end{equation}
The proportionality constant has to be fixed by requiring that 
$\tilde{r}(r) \to r$ when $r \to \infty$.
Summarizing, the mapping between the metric functions in the line elements 
\eqref{quasiisometric} and \eqref{polarmetric} is 
given by
\begin{align}
 \alpha(\tilde{r})&=\log{\Bigl(\frac{r(\tilde{r})}{\tilde{r}}\Bigr)}\ ,\nonumber\\ 
 \gamma(\tilde{r})&=\rho(\tilde{r})+2\alpha(\tilde{r})\ ,\nonumber\\
\rho(\tilde{r})&=\frac{1}{2}v(r(\tilde{r}))-\alpha(\tilde{r})\ .
\end{align}

\section{Comparison with previous results}\label{app:comparions}
We have tested the validity of our approach by comparing 
the numerical values obtained for the multipole 
moments of rotating BS, with previous results known 
in literature. 

\begin{figure*}[htp]
    \centering
    \includegraphics[width=0.49\textwidth]{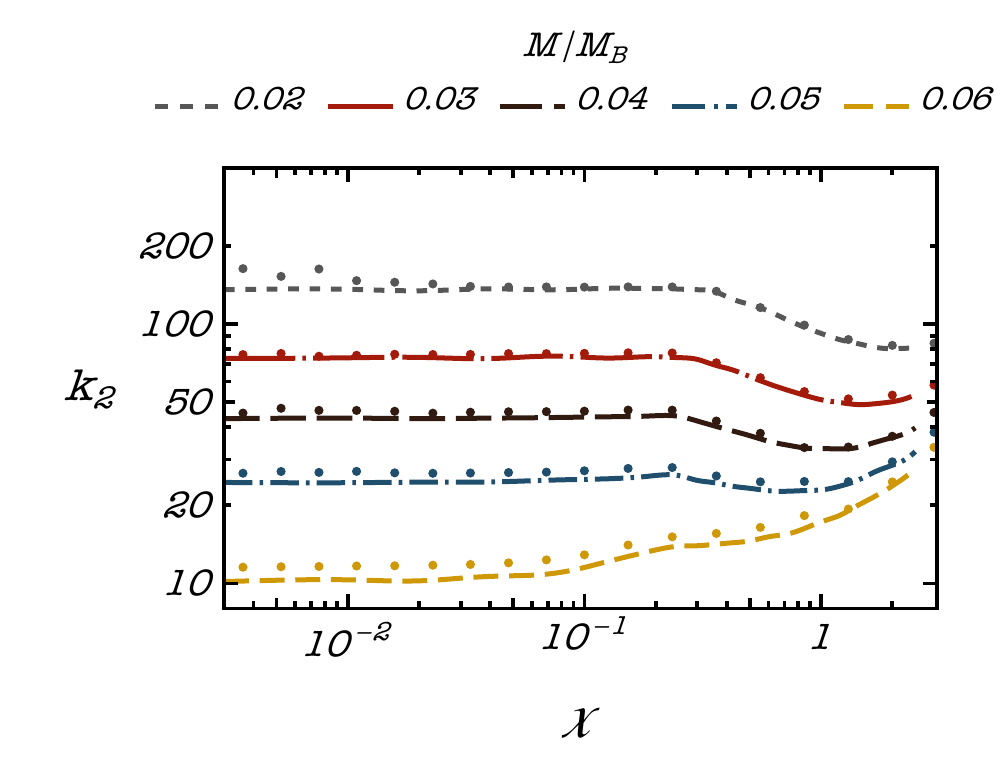}
    \includegraphics[width=0.49\textwidth]{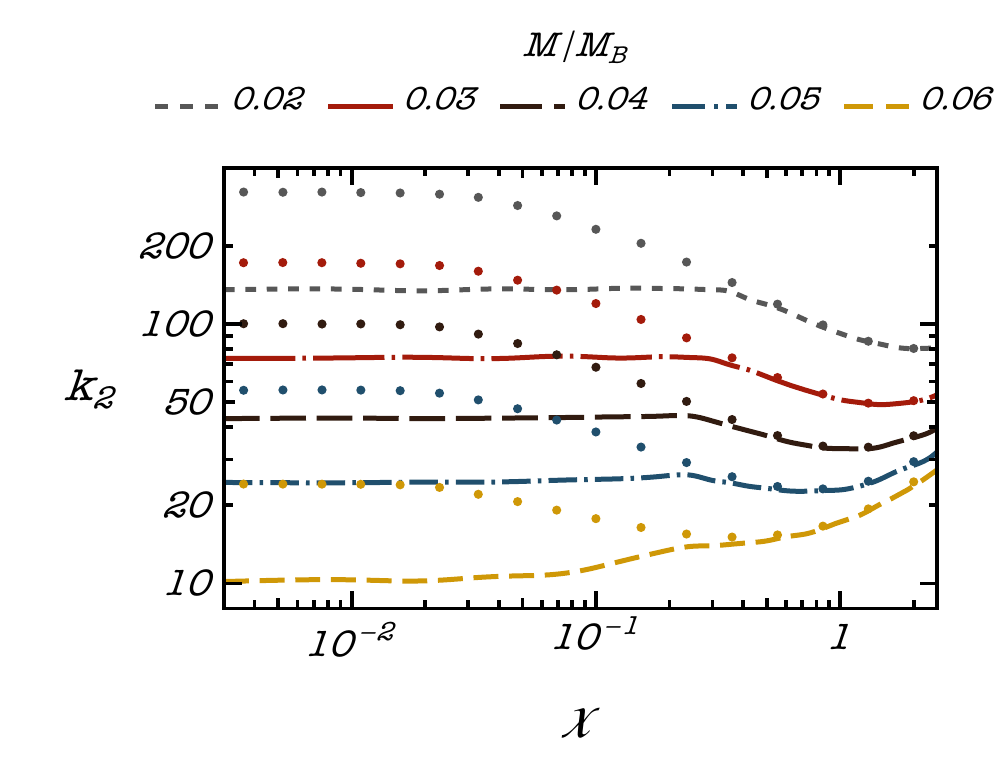}
    \caption{Left: Reduced quadrupole moment $\kappa_2$ as a function 
    of the dimensionless spin $\chi$. Data points  
    correspond to results obtained in this work setting the 
    grid for the numerical integration of the field's equation 
    to $n_q \times n_\mu=1600 \times 160$. Dashed curves  
    refer to fit of data computed with the same set up 
    in~\cite{ryan_spinning_1997}.
    Right: same as the left panel but with higher angular resolution, $n_q \times n_\mu=1600 \times 20000$. The low spin constant value is a factor of $\sim 2$ larger than in~\cite{ryan_spinning_1997} (dashed lines) , while there is a very good agreement for $\chi \gtrsim 0.3$.
    }
    \label{fig:Ryan_cfr}
\end{figure*}

The left panel of Fig.~\ref{fig:Ryan_cfr} shows the reduced quadrupole 
moment $\kappa_2$ computed with our code, for five BS 
families with different masses, as a function of the 
spin $\chi$, compared against the values 
obtained in~\cite{ryan_spinning_1997}. 
Each point represents a different BS solution 
derived by solving field's equation on a 
grid $n_q \times n_\mu=1600 \times 160$, which is 
the same adopted in~\cite{ryan_spinning_1997}. 
Dashed lines correspond to data extracted from Fig.~4 of~\cite{ryan_spinning_1997}. 
The values of the reduced quadrupole agree remarkably 
well on a wide range of spins, with an average relative 
discrepancy smaller than $7\%$ for all 
the considered BS masses. 

However, as discussed in Sec.~\ref{Sec:scheme}-\ref{sec:results}, 
calculations of the multipole moments are sensitive to the 
choice of the angular spacing $n_\mu$ for small values of 
the spin. By increasing $n_\mu$ we find indeed that 
the values of $\kappa_2$ (and $\sigma_3$) start deviating 
from those obtained in~\cite{ryan_spinning_1997}.

In the right panel of Fig.~\ref{fig:Ryan_cfr} we show the reduced quadrupole 
computed using a $n_q\times n_\mu=1600 \times 20000$ grid, 
again compared with data produced in~\cite{ryan_spinning_1997}. 
While at high spins, the agreement between the two sets 
of results still hold, for rotating BSs with 
$\chi\lesssim0.1$ our values are in general 
larger by a factor $\sim2$ than those calculated by Ryan. 
We also note that for $M=0.06M_B$, the reduced quadrupole
obtained with increased accuracy features an overall change 
in the behavior of $\kappa_2$ as a function of $\chi$. 
As explained in Sec.~\ref{sec:results} we have checked that 
our results saturate for large enough $n_\mu$. Indeed, 
doubling the grid resolution from $n_\mu=20000$ to $n_\mu=40000$ leads 
to variation in the multipole moments smaller than $5\%$ for $\chi \lesssim 0.1$ and even less for larger spins. 

\bibliography{BSMultipoles}

\begin{thebibliography}{58}%
\makeatletter
\providecommand \@ifxundefined [1]{%
 \@ifx{#1\undefined}
}%
\providecommand \@ifnum [1]{%
 \ifnum #1\expandafter \@firstoftwo
 \else \expandafter \@secondoftwo
 \fi
}%
\providecommand \@ifx [1]{%
 \ifx #1\expandafter \@firstoftwo
 \else \expandafter \@secondoftwo
 \fi
}%
\providecommand \natexlab [1]{#1}%
\providecommand \enquote  [1]{``#1''}%
\providecommand \bibnamefont  [1]{#1}%
\providecommand \bibfnamefont [1]{#1}%
\providecommand \citenamefont [1]{#1}%
\providecommand \href@noop [0]{\@secondoftwo}%
\providecommand \href [0]{\begingroup \@sanitize@url \@href}%
\providecommand \@href[1]{\@@startlink{#1}\@@href}%
\providecommand \@@href[1]{\endgroup#1\@@endlink}%
\providecommand \@sanitize@url [0]{\catcode `\\12\catcode `\$12\catcode
  `\&12\catcode `\#12\catcode `\^12\catcode `\_12\catcode `\%12\relax}%
\providecommand \@@startlink[1]{}%
\providecommand \@@endlink[0]{}%
\providecommand \url  [0]{\begingroup\@sanitize@url \@url }%
\providecommand \@url [1]{\endgroup\@href {#1}{\urlprefix }}%
\providecommand \urlprefix  [0]{URL }%
\providecommand \Eprint [0]{\href }%
\providecommand \doibase [0]{https://doi.org/}%
\providecommand \selectlanguage [0]{\@gobble}%
\providecommand \bibinfo  [0]{\@secondoftwo}%
\providecommand \bibfield  [0]{\@secondoftwo}%
\providecommand \translation [1]{[#1]}%
\providecommand \BibitemOpen [0]{}%
\providecommand \bibitemStop [0]{}%
\providecommand \bibitemNoStop [0]{.\EOS\space}%
\providecommand \EOS [0]{\spacefactor3000\relax}%
\providecommand \BibitemShut  [1]{\csname bibitem#1\endcsname}%
\let\auto@bib@innerbib\@empty
\bibitem [{\citenamefont {Barack}\ \emph {et~al.}(2019)\citenamefont {Barack}
  \emph {et~al.}}]{Barack:2018yly}%
  \BibitemOpen
  \bibfield  {author} {\bibinfo {author} {\bibfnamefont {L.}~\bibnamefont
  {Barack}} \emph {et~al.},\ }\href {https://doi.org/10.1088/1361-6382/ab0587}
  {\bibfield  {journal} {\bibinfo  {journal} {Class. Quant. Grav.}\ }\textbf
  {\bibinfo {volume} {36}},\ \bibinfo {pages} {143001} (\bibinfo {year}
  {2019})},\ \Eprint {https://arxiv.org/abs/1806.05195} {arXiv:1806.05195
  [gr-qc]} \BibitemShut {NoStop}%
\bibitem [{\citenamefont {Cardoso}\ and\ \citenamefont
  {Pani}(2019)}]{Cardoso:2019rvt}%
  \BibitemOpen
  \bibfield  {author} {\bibinfo {author} {\bibfnamefont {V.}~\bibnamefont
  {Cardoso}}\ and\ \bibinfo {author} {\bibfnamefont {P.}~\bibnamefont {Pani}},\
  }\href {https://doi.org/10.1007/s41114-019-0020-4} {\bibfield  {journal}
  {\bibinfo  {journal} {Living Rev. Rel.}\ }\textbf {\bibinfo {volume} {22}},\
  \bibinfo {pages} {4} (\bibinfo {year} {2019})},\ \Eprint
  {https://arxiv.org/abs/1904.05363} {arXiv:1904.05363 [gr-qc]} \BibitemShut
  {NoStop}%
\bibitem [{\citenamefont {Abbott}\ \emph {et~al.}(2021)\citenamefont {Abbott}
  \emph {et~al.}}]{LIGOScientific:2021sio}%
  \BibitemOpen
  \bibfield  {author} {\bibinfo {author} {\bibfnamefont {R.}~\bibnamefont
  {Abbott}} \emph {et~al.} (\bibinfo {collaboration} {LIGO Scientific, VIRGO,
  KAGRA}),\ }\href@noop {} {\  (\bibinfo {year} {2021})},\ \Eprint
  {https://arxiv.org/abs/2112.06861} {arXiv:2112.06861 [gr-qc]} \BibitemShut
  {NoStop}%
\bibitem [{\citenamefont {Maggio}\ \emph {et~al.}(2021)\citenamefont {Maggio},
  \citenamefont {Pani},\ and\ \citenamefont {Raposo}}]{Maggio:2021ans}%
  \BibitemOpen
  \bibfield  {author} {\bibinfo {author} {\bibfnamefont {E.}~\bibnamefont
  {Maggio}}, \bibinfo {author} {\bibfnamefont {P.}~\bibnamefont {Pani}},\ and\
  \bibinfo {author} {\bibfnamefont {G.}~\bibnamefont {Raposo}},\ }\href@noop {}
  {\  (\bibinfo {year} {2021})},\ \Eprint {https://arxiv.org/abs/2105.06410}
  {arXiv:2105.06410 [gr-qc]} \BibitemShut {NoStop}%
\bibitem [{\citenamefont {Giudice}\ \emph {et~al.}(2016)\citenamefont
  {Giudice}, \citenamefont {McCullough},\ and\ \citenamefont
  {Urbano}}]{Giudice:2016zpa}%
  \BibitemOpen
  \bibfield  {author} {\bibinfo {author} {\bibfnamefont {G.~F.}\ \bibnamefont
  {Giudice}}, \bibinfo {author} {\bibfnamefont {M.}~\bibnamefont
  {McCullough}},\ and\ \bibinfo {author} {\bibfnamefont {A.}~\bibnamefont
  {Urbano}},\ }\href {https://doi.org/10.1088/1475-7516/2016/10/001} {\bibfield
   {journal} {\bibinfo  {journal} {JCAP}\ }\textbf {\bibinfo {volume} {10}},\
  \bibinfo {pages} {001}},\ \Eprint {https://arxiv.org/abs/1605.01209}
  {arXiv:1605.01209 [hep-ph]} \BibitemShut {NoStop}%
\bibitem [{\citenamefont {Pacilio}\ \emph {et~al.}(2020)\citenamefont
  {Pacilio}, \citenamefont {Vaglio}, \citenamefont {Maselli},\ and\
  \citenamefont {Pani}}]{pacilio_gravitational-wave_2020}%
  \BibitemOpen
  \bibfield  {author} {\bibinfo {author} {\bibfnamefont {C.}~\bibnamefont
  {Pacilio}}, \bibinfo {author} {\bibfnamefont {M.}~\bibnamefont {Vaglio}},
  \bibinfo {author} {\bibfnamefont {A.}~\bibnamefont {Maselli}},\ and\ \bibinfo
  {author} {\bibfnamefont {P.}~\bibnamefont {Pani}},\ }\href
  {https://doi.org/10.1103/PhysRevD.102.083002} {\bibfield  {journal} {\bibinfo
   {journal} {Physical Review D}\ }\textbf {\bibinfo {volume} {102}},\ \bibinfo
  {pages} {083002} (\bibinfo {year} {2020})}\BibitemShut {NoStop}%
\bibitem [{\citenamefont {Bustillo}\ \emph {et~al.}(2021)\citenamefont
  {Bustillo}, \citenamefont {Sanchis-Gual}, \citenamefont {Torres-Forn\'e},
  \citenamefont {Font}, \citenamefont {Vajpeyi}, \citenamefont {Smith},
  \citenamefont {Herdeiro}, \citenamefont {Radu},\ and\ \citenamefont
  {Leong}}]{Bustillo:2020syj}%
  \BibitemOpen
  \bibfield  {author} {\bibinfo {author} {\bibfnamefont {J.~C.}\ \bibnamefont
  {Bustillo}}, \bibinfo {author} {\bibfnamefont {N.}~\bibnamefont
  {Sanchis-Gual}}, \bibinfo {author} {\bibfnamefont {A.}~\bibnamefont
  {Torres-Forn\'e}}, \bibinfo {author} {\bibfnamefont {J.~A.}\ \bibnamefont
  {Font}}, \bibinfo {author} {\bibfnamefont {A.}~\bibnamefont {Vajpeyi}},
  \bibinfo {author} {\bibfnamefont {R.}~\bibnamefont {Smith}}, \bibinfo
  {author} {\bibfnamefont {C.}~\bibnamefont {Herdeiro}}, \bibinfo {author}
  {\bibfnamefont {E.}~\bibnamefont {Radu}},\ and\ \bibinfo {author}
  {\bibfnamefont {S.~H.~W.}\ \bibnamefont {Leong}},\ }\href
  {https://doi.org/10.1103/PhysRevLett.126.081101} {\bibfield  {journal}
  {\bibinfo  {journal} {Phys. Rev. Lett.}\ }\textbf {\bibinfo {volume} {126}},\
  \bibinfo {pages} {081101} (\bibinfo {year} {2021})},\ \Eprint
  {https://arxiv.org/abs/2009.05376} {arXiv:2009.05376 [gr-qc]} \BibitemShut
  {NoStop}%
\bibitem [{\citenamefont {Abbott}\ \emph
  {et~al.}(2020{\natexlab{a}})\citenamefont {Abbott} \emph
  {et~al.}}]{LIGOScientific:2020ufj}%
  \BibitemOpen
  \bibfield  {author} {\bibinfo {author} {\bibfnamefont {R.}~\bibnamefont
  {Abbott}} \emph {et~al.} (\bibinfo {collaboration} {LIGO Scientific,
  Virgo}),\ }\href {https://doi.org/10.3847/2041-8213/aba493} {\bibfield
  {journal} {\bibinfo  {journal} {Astrophys. J. Lett.}\ }\textbf {\bibinfo
  {volume} {900}},\ \bibinfo {pages} {L13} (\bibinfo {year}
  {2020}{\natexlab{a}})},\ \Eprint {https://arxiv.org/abs/2009.01190}
  {arXiv:2009.01190 [astro-ph.HE]} \BibitemShut {NoStop}%
\bibitem [{\citenamefont {Abbott}\ \emph
  {et~al.}(2020{\natexlab{b}})\citenamefont {Abbott} \emph
  {et~al.}}]{LIGOScientific:2020iuh}%
  \BibitemOpen
  \bibfield  {author} {\bibinfo {author} {\bibfnamefont {R.}~\bibnamefont
  {Abbott}} \emph {et~al.} (\bibinfo {collaboration} {LIGO Scientific,
  Virgo}),\ }\href {https://doi.org/10.1103/PhysRevLett.125.101102} {\bibfield
  {journal} {\bibinfo  {journal} {Phys. Rev. Lett.}\ }\textbf {\bibinfo
  {volume} {125}},\ \bibinfo {pages} {101102} (\bibinfo {year}
  {2020}{\natexlab{b}})},\ \Eprint {https://arxiv.org/abs/2009.01075}
  {arXiv:2009.01075 [gr-qc]} \BibitemShut {NoStop}%
\bibitem [{\citenamefont {Kaup}(1968)}]{Kaup:1968zz}%
  \BibitemOpen
  \bibfield  {author} {\bibinfo {author} {\bibfnamefont {D.~J.}\ \bibnamefont
  {Kaup}},\ }\href {https://doi.org/10.1103/PhysRev.172.1331} {\bibfield
  {journal} {\bibinfo  {journal} {Phys. Rev.}\ }\textbf {\bibinfo {volume}
  {172}},\ \bibinfo {pages} {1331} (\bibinfo {year} {1968})}\BibitemShut
  {NoStop}%
\bibitem [{\citenamefont {Ruffini}\ and\ \citenamefont
  {Bonazzola}(1969)}]{Ruffini:1969qy}%
  \BibitemOpen
  \bibfield  {author} {\bibinfo {author} {\bibfnamefont {R.}~\bibnamefont
  {Ruffini}}\ and\ \bibinfo {author} {\bibfnamefont {S.}~\bibnamefont
  {Bonazzola}},\ }\href {https://doi.org/10.1103/PhysRev.187.1767} {\bibfield
  {journal} {\bibinfo  {journal} {Phys. Rev.}\ }\textbf {\bibinfo {volume}
  {187}},\ \bibinfo {pages} {1767} (\bibinfo {year} {1969})}\BibitemShut
  {NoStop}%
\bibitem [{\citenamefont {Colpi}\ \emph {et~al.}(1986)\citenamefont {Colpi},
  \citenamefont {Shapiro},\ and\ \citenamefont {Wasserman}}]{colpi_boson_1986}%
  \BibitemOpen
  \bibfield  {author} {\bibinfo {author} {\bibfnamefont {M.}~\bibnamefont
  {Colpi}}, \bibinfo {author} {\bibfnamefont {S.~L.}\ \bibnamefont {Shapiro}},\
  and\ \bibinfo {author} {\bibfnamefont {I.}~\bibnamefont {Wasserman}},\ }\href
  {https://doi.org/10.1103/PhysRevLett.57.2485} {\bibfield  {journal} {\bibinfo
   {journal} {Physical Review Letters}\ }\textbf {\bibinfo {volume} {57}},\
  \bibinfo {pages} {2485} (\bibinfo {year} {1986})}\BibitemShut {NoStop}%
\bibitem [{\citenamefont {Brito}\ \emph {et~al.}(2016)\citenamefont {Brito},
  \citenamefont {Cardoso}, \citenamefont {Herdeiro},\ and\ \citenamefont
  {Radu}}]{Brito:2015pxa}%
  \BibitemOpen
  \bibfield  {author} {\bibinfo {author} {\bibfnamefont {R.}~\bibnamefont
  {Brito}}, \bibinfo {author} {\bibfnamefont {V.}~\bibnamefont {Cardoso}},
  \bibinfo {author} {\bibfnamefont {C.~A.~R.}\ \bibnamefont {Herdeiro}},\ and\
  \bibinfo {author} {\bibfnamefont {E.}~\bibnamefont {Radu}},\ }\href
  {https://doi.org/10.1016/j.physletb.2015.11.051} {\bibfield  {journal}
  {\bibinfo  {journal} {Phys. Lett. B}\ }\textbf {\bibinfo {volume} {752}},\
  \bibinfo {pages} {291} (\bibinfo {year} {2016})},\ \Eprint
  {https://arxiv.org/abs/1508.05395} {arXiv:1508.05395 [gr-qc]} \BibitemShut
  {NoStop}%
\bibitem [{\citenamefont {Jetzer}(1992)}]{Jetzer:1991jr}%
  \BibitemOpen
  \bibfield  {author} {\bibinfo {author} {\bibfnamefont {P.}~\bibnamefont
  {Jetzer}},\ }\href {https://doi.org/10.1016/0370-1573(92)90123-H} {\bibfield
  {journal} {\bibinfo  {journal} {Phys. Rept.}\ }\textbf {\bibinfo {volume}
  {220}},\ \bibinfo {pages} {163} (\bibinfo {year} {1992})}\BibitemShut
  {NoStop}%
\bibitem [{\citenamefont {Liebling}\ and\ \citenamefont
  {Palenzuela}(2017)}]{liebling_dynamical_2017}%
  \BibitemOpen
  \bibfield  {author} {\bibinfo {author} {\bibfnamefont {S.~L.}\ \bibnamefont
  {Liebling}}\ and\ \bibinfo {author} {\bibfnamefont {C.}~\bibnamefont
  {Palenzuela}},\ }\href {https://doi.org/10.1007/s41114-017-0007-y} {\bibfield
   {journal} {\bibinfo  {journal} {Living Reviews in Relativity}\ }\textbf
  {\bibinfo {volume} {20}},\ \bibinfo {pages} {5} (\bibinfo {year}
  {2017})}\BibitemShut {NoStop}%
\bibitem [{\citenamefont {Palenzuela}\ \emph {et~al.}(2008)\citenamefont
  {Palenzuela}, \citenamefont {Lehner},\ and\ \citenamefont
  {Liebling}}]{Palenzuela:2007dm}%
  \BibitemOpen
  \bibfield  {author} {\bibinfo {author} {\bibfnamefont {C.}~\bibnamefont
  {Palenzuela}}, \bibinfo {author} {\bibfnamefont {L.}~\bibnamefont {Lehner}},\
  and\ \bibinfo {author} {\bibfnamefont {S.~L.}\ \bibnamefont {Liebling}},\
  }\href {https://doi.org/10.1103/PhysRevD.77.044036} {\bibfield  {journal}
  {\bibinfo  {journal} {Phys. Rev. D}\ }\textbf {\bibinfo {volume} {77}},\
  \bibinfo {pages} {044036} (\bibinfo {year} {2008})},\ \Eprint
  {https://arxiv.org/abs/0706.2435} {arXiv:0706.2435 [gr-qc]} \BibitemShut
  {NoStop}%
\bibitem [{\citenamefont {Palenzuela}\ \emph {et~al.}(2017)\citenamefont
  {Palenzuela}, \citenamefont {Pani}, \citenamefont {Bezares}, \citenamefont
  {Cardoso}, \citenamefont {Lehner},\ and\ \citenamefont
  {Liebling}}]{Palenzuela:2017kcg}%
  \BibitemOpen
  \bibfield  {author} {\bibinfo {author} {\bibfnamefont {C.}~\bibnamefont
  {Palenzuela}}, \bibinfo {author} {\bibfnamefont {P.}~\bibnamefont {Pani}},
  \bibinfo {author} {\bibfnamefont {M.}~\bibnamefont {Bezares}}, \bibinfo
  {author} {\bibfnamefont {V.}~\bibnamefont {Cardoso}}, \bibinfo {author}
  {\bibfnamefont {L.}~\bibnamefont {Lehner}},\ and\ \bibinfo {author}
  {\bibfnamefont {S.}~\bibnamefont {Liebling}},\ }\href
  {https://doi.org/10.1103/PhysRevD.96.104058} {\bibfield  {journal} {\bibinfo
  {journal} {Phys. Rev. D}\ }\textbf {\bibinfo {volume} {96}},\ \bibinfo
  {pages} {104058} (\bibinfo {year} {2017})},\ \Eprint
  {https://arxiv.org/abs/1710.09432} {arXiv:1710.09432 [gr-qc]} \BibitemShut
  {NoStop}%
\bibitem [{\citenamefont {Bezares}\ and\ \citenamefont
  {Palenzuela}(2018)}]{Bezares:2018qwa}%
  \BibitemOpen
  \bibfield  {author} {\bibinfo {author} {\bibfnamefont {M.}~\bibnamefont
  {Bezares}}\ and\ \bibinfo {author} {\bibfnamefont {C.}~\bibnamefont
  {Palenzuela}},\ }\href {https://doi.org/10.1088/1361-6382/aae87c} {\bibfield
  {journal} {\bibinfo  {journal} {Class. Quant. Grav.}\ }\textbf {\bibinfo
  {volume} {35}},\ \bibinfo {pages} {234002} (\bibinfo {year} {2018})},\
  \Eprint {https://arxiv.org/abs/1808.10732} {arXiv:1808.10732 [gr-qc]}
  \BibitemShut {NoStop}%
\bibitem [{\citenamefont {Bezares}\ \emph {et~al.}(2022)\citenamefont
  {Bezares}, \citenamefont {Bo\v{s}kovi\'c}, \citenamefont {Liebling},
  \citenamefont {Palenzuela}, \citenamefont {Pani},\ and\ \citenamefont
  {Barausse}}]{Bezares:2022obu}%
  \BibitemOpen
  \bibfield  {author} {\bibinfo {author} {\bibfnamefont {M.}~\bibnamefont
  {Bezares}}, \bibinfo {author} {\bibfnamefont {M.}~\bibnamefont
  {Bo\v{s}kovi\'c}}, \bibinfo {author} {\bibfnamefont {S.}~\bibnamefont
  {Liebling}}, \bibinfo {author} {\bibfnamefont {C.}~\bibnamefont
  {Palenzuela}}, \bibinfo {author} {\bibfnamefont {P.}~\bibnamefont {Pani}},\
  and\ \bibinfo {author} {\bibfnamefont {E.}~\bibnamefont {Barausse}},\
  }\href@noop {} {\  (\bibinfo {year} {2022})},\ \Eprint
  {https://arxiv.org/abs/2201.06113} {arXiv:2201.06113 [gr-qc]} \BibitemShut
  {NoStop}%
\bibitem [{\citenamefont {Sanchis-Gual}\ \emph {et~al.}(2019)\citenamefont
  {Sanchis-Gual}, \citenamefont {Di~Giovanni}, \citenamefont {Zilh\~ao},
  \citenamefont {Herdeiro}, \citenamefont {Cerd\'a-Dur\'an}, \citenamefont
  {Font},\ and\ \citenamefont {Radu}}]{Sanchis-Gual:2019ljs}%
  \BibitemOpen
  \bibfield  {author} {\bibinfo {author} {\bibfnamefont {N.}~\bibnamefont
  {Sanchis-Gual}}, \bibinfo {author} {\bibfnamefont {F.}~\bibnamefont
  {Di~Giovanni}}, \bibinfo {author} {\bibfnamefont {M.}~\bibnamefont
  {Zilh\~ao}}, \bibinfo {author} {\bibfnamefont {C.}~\bibnamefont {Herdeiro}},
  \bibinfo {author} {\bibfnamefont {P.}~\bibnamefont {Cerd\'a-Dur\'an}},
  \bibinfo {author} {\bibfnamefont {J.~A.}\ \bibnamefont {Font}},\ and\
  \bibinfo {author} {\bibfnamefont {E.}~\bibnamefont {Radu}},\ }\href
  {https://doi.org/10.1103/PhysRevLett.123.221101} {\bibfield  {journal}
  {\bibinfo  {journal} {Phys. Rev. Lett.}\ }\textbf {\bibinfo {volume} {123}},\
  \bibinfo {pages} {221101} (\bibinfo {year} {2019})},\ \Eprint
  {https://arxiv.org/abs/1907.12565} {arXiv:1907.12565 [gr-qc]} \BibitemShut
  {NoStop}%
\bibitem [{\citenamefont {Di~Giovanni}\ \emph {et~al.}(2020)\citenamefont
  {Di~Giovanni}, \citenamefont {Sanchis-Gual}, \citenamefont {Cerd\'a-Dur\'an},
  \citenamefont {Zilh\~ao}, \citenamefont {Herdeiro}, \citenamefont {Font},\
  and\ \citenamefont {Radu}}]{DiGiovanni:2020ror}%
  \BibitemOpen
  \bibfield  {author} {\bibinfo {author} {\bibfnamefont {F.}~\bibnamefont
  {Di~Giovanni}}, \bibinfo {author} {\bibfnamefont {N.}~\bibnamefont
  {Sanchis-Gual}}, \bibinfo {author} {\bibfnamefont {P.}~\bibnamefont
  {Cerd\'a-Dur\'an}}, \bibinfo {author} {\bibfnamefont {M.}~\bibnamefont
  {Zilh\~ao}}, \bibinfo {author} {\bibfnamefont {C.}~\bibnamefont {Herdeiro}},
  \bibinfo {author} {\bibfnamefont {J.~A.}\ \bibnamefont {Font}},\ and\
  \bibinfo {author} {\bibfnamefont {E.}~\bibnamefont {Radu}},\ }\href
  {https://doi.org/10.1103/PhysRevD.102.124009} {\bibfield  {journal} {\bibinfo
   {journal} {Phys. Rev. D}\ }\textbf {\bibinfo {volume} {102}},\ \bibinfo
  {pages} {124009} (\bibinfo {year} {2020})},\ \Eprint
  {https://arxiv.org/abs/2010.05845} {arXiv:2010.05845 [gr-qc]} \BibitemShut
  {NoStop}%
\bibitem [{\citenamefont {Siemonsen}\ and\ \citenamefont
  {East}(2021)}]{siemonsen_stability_2021}%
  \BibitemOpen
  \bibfield  {author} {\bibinfo {author} {\bibfnamefont {N.}~\bibnamefont
  {Siemonsen}}\ and\ \bibinfo {author} {\bibfnamefont {W.~E.}\ \bibnamefont
  {East}},\ }\href {https://doi.org/10.1103/PhysRevD.103.044022} {\bibfield
  {journal} {\bibinfo  {journal} {Physical Review D}\ }\textbf {\bibinfo
  {volume} {103}},\ \bibinfo {pages} {044022} (\bibinfo {year}
  {2021})}\BibitemShut {NoStop}%
\bibitem [{\citenamefont {Poisson}\ and\ \citenamefont
  {Will}(2014)}]{PoissonWill}%
  \BibitemOpen
  \bibfield  {author} {\bibinfo {author} {\bibfnamefont {E.}~\bibnamefont
  {Poisson}}\ and\ \bibinfo {author} {\bibfnamefont {C.}~\bibnamefont {Will}},\
  }\href {https://books.google.it/books?id=PZ5cAwAAQBAJ} {\emph {\bibinfo
  {title} {Gravity: Newtonian, Post-Newtonian, Relativistic}}}\ (\bibinfo
  {publisher} {Cambridge University Press},\ \bibinfo {year}
  {2014})\BibitemShut {NoStop}%
\bibitem [{\citenamefont {Hansen}(1974{\natexlab{a}})}]{Hansen:1974zz}%
  \BibitemOpen
  \bibfield  {author} {\bibinfo {author} {\bibfnamefont {R.~O.}\ \bibnamefont
  {Hansen}},\ }\href {https://doi.org/10.1063/1.1666501} {\bibfield  {journal}
  {\bibinfo  {journal} {J. Math. Phys.}\ }\textbf {\bibinfo {volume} {15}},\
  \bibinfo {pages} {46} (\bibinfo {year} {1974}{\natexlab{a}})}\BibitemShut
  {NoStop}%
\bibitem [{\citenamefont {Geroch}(1970)}]{Geroch:1970cd}%
  \BibitemOpen
  \bibfield  {author} {\bibinfo {author} {\bibfnamefont {R.~P.}\ \bibnamefont
  {Geroch}},\ }\href {https://doi.org/10.1063/1.1665427} {\bibfield  {journal}
  {\bibinfo  {journal} {J. Math. Phys.}\ }\textbf {\bibinfo {volume} {11}},\
  \bibinfo {pages} {2580} (\bibinfo {year} {1970})}\BibitemShut {NoStop}%
\bibitem [{\citenamefont {Bianchi}\ \emph {et~al.}(2020)\citenamefont
  {Bianchi}, \citenamefont {Consoli}, \citenamefont {Grillo}, \citenamefont
  {Morales}, \citenamefont {Pani},\ and\ \citenamefont
  {Raposo}}]{Bianchi:2020bxa}%
  \BibitemOpen
  \bibfield  {author} {\bibinfo {author} {\bibfnamefont {M.}~\bibnamefont
  {Bianchi}}, \bibinfo {author} {\bibfnamefont {D.}~\bibnamefont {Consoli}},
  \bibinfo {author} {\bibfnamefont {A.}~\bibnamefont {Grillo}}, \bibinfo
  {author} {\bibfnamefont {J.~F.}\ \bibnamefont {Morales}}, \bibinfo {author}
  {\bibfnamefont {P.}~\bibnamefont {Pani}},\ and\ \bibinfo {author}
  {\bibfnamefont {G.}~\bibnamefont {Raposo}},\ }\href
  {https://doi.org/10.1103/PhysRevLett.125.221601} {\bibfield  {journal}
  {\bibinfo  {journal} {Phys. Rev. Lett.}\ }\textbf {\bibinfo {volume} {125}},\
  \bibinfo {pages} {221601} (\bibinfo {year} {2020})},\ \Eprint
  {https://arxiv.org/abs/2007.01743} {arXiv:2007.01743 [hep-th]} \BibitemShut
  {NoStop}%
\bibitem [{\citenamefont {Bianchi}\ \emph {et~al.}(2021)\citenamefont
  {Bianchi}, \citenamefont {Consoli}, \citenamefont {Grillo}, \citenamefont
  {Morales}, \citenamefont {Pani},\ and\ \citenamefont
  {Raposo}}]{Bianchi:2020miz}%
  \BibitemOpen
  \bibfield  {author} {\bibinfo {author} {\bibfnamefont {M.}~\bibnamefont
  {Bianchi}}, \bibinfo {author} {\bibfnamefont {D.}~\bibnamefont {Consoli}},
  \bibinfo {author} {\bibfnamefont {A.}~\bibnamefont {Grillo}}, \bibinfo
  {author} {\bibfnamefont {J.~F.}\ \bibnamefont {Morales}}, \bibinfo {author}
  {\bibfnamefont {P.}~\bibnamefont {Pani}},\ and\ \bibinfo {author}
  {\bibfnamefont {G.}~\bibnamefont {Raposo}},\ }\href
  {https://doi.org/10.1007/JHEP01(2021)003} {\bibfield  {journal} {\bibinfo
  {journal} {JHEP}\ }\textbf {\bibinfo {volume} {01}},\ \bibinfo {pages}
  {003}},\ \Eprint {https://arxiv.org/abs/2008.01445} {arXiv:2008.01445
  [hep-th]} \BibitemShut {NoStop}%
\bibitem [{\citenamefont {Fransen}\ and\ \citenamefont
  {Mayerson}(2022)}]{Fransen:2022jtw}%
  \BibitemOpen
  \bibfield  {author} {\bibinfo {author} {\bibfnamefont {K.}~\bibnamefont
  {Fransen}}\ and\ \bibinfo {author} {\bibfnamefont {D.~R.}\ \bibnamefont
  {Mayerson}},\ }\href@noop {} {\  (\bibinfo {year} {2022})},\ \Eprint
  {https://arxiv.org/abs/2201.03569} {arXiv:2201.03569 [gr-qc]} \BibitemShut
  {NoStop}%
\bibitem [{\citenamefont {Loutrel}\ \emph {et~al.}(2022)\citenamefont
  {Loutrel}, \citenamefont {Brito}, \citenamefont {Maselli},\ and\
  \citenamefont {Pani}}]{Loutrel:2022ant}%
  \BibitemOpen
  \bibfield  {author} {\bibinfo {author} {\bibfnamefont {N.}~\bibnamefont
  {Loutrel}}, \bibinfo {author} {\bibfnamefont {R.}~\bibnamefont {Brito}},
  \bibinfo {author} {\bibfnamefont {A.}~\bibnamefont {Maselli}},\ and\ \bibinfo
  {author} {\bibfnamefont {P.}~\bibnamefont {Pani}},\ }\href@noop {} {\
  (\bibinfo {year} {2022})},\ \Eprint {https://arxiv.org/abs/2203.01725}
  {arXiv:2203.01725 [gr-qc]} \BibitemShut {NoStop}%
\bibitem [{\citenamefont {Psaltis}(2008)}]{Psaltis:2008bb}%
  \BibitemOpen
  \bibfield  {author} {\bibinfo {author} {\bibfnamefont {D.}~\bibnamefont
  {Psaltis}},\ }\href {https://doi.org/10.12942/lrr-2008-9} {\bibfield
  {journal} {\bibinfo  {journal} {Living Rev. Rel.}\ }\textbf {\bibinfo
  {volume} {11}},\ \bibinfo {pages} {9} (\bibinfo {year} {2008})},\ \Eprint
  {https://arxiv.org/abs/0806.1531} {arXiv:0806.1531 [astro-ph]} \BibitemShut
  {NoStop}%
\bibitem [{\citenamefont {Gair}\ \emph {et~al.}(2013)\citenamefont {Gair},
  \citenamefont {Vallisneri}, \citenamefont {Larson},\ and\ \citenamefont
  {Baker}}]{Gair:2012nm}%
  \BibitemOpen
  \bibfield  {author} {\bibinfo {author} {\bibfnamefont {J.~R.}\ \bibnamefont
  {Gair}}, \bibinfo {author} {\bibfnamefont {M.}~\bibnamefont {Vallisneri}},
  \bibinfo {author} {\bibfnamefont {S.~L.}\ \bibnamefont {Larson}},\ and\
  \bibinfo {author} {\bibfnamefont {J.~G.}\ \bibnamefont {Baker}},\ }\href
  {https://doi.org/10.12942/lrr-2013-7} {\bibfield  {journal} {\bibinfo
  {journal} {Living Rev. Rel.}\ }\textbf {\bibinfo {volume} {16}},\ \bibinfo
  {pages} {7} (\bibinfo {year} {2013})},\ \Eprint
  {https://arxiv.org/abs/1212.5575} {arXiv:1212.5575 [gr-qc]} \BibitemShut
  {NoStop}%
\bibitem [{\citenamefont {Yunes}\ and\ \citenamefont
  {Siemens}(2013)}]{Yunes:2013dva}%
  \BibitemOpen
  \bibfield  {author} {\bibinfo {author} {\bibfnamefont {N.}~\bibnamefont
  {Yunes}}\ and\ \bibinfo {author} {\bibfnamefont {X.}~\bibnamefont
  {Siemens}},\ }\href {https://doi.org/10.12942/lrr-2013-9} {\bibfield
  {journal} {\bibinfo  {journal} {Living Rev. Rel.}\ }\textbf {\bibinfo
  {volume} {16}},\ \bibinfo {pages} {9} (\bibinfo {year} {2013})},\ \Eprint
  {https://arxiv.org/abs/1304.3473} {arXiv:1304.3473 [gr-qc]} \BibitemShut
  {NoStop}%
\bibitem [{\citenamefont {Berti}\ \emph {et~al.}(2015)\citenamefont {Berti}
  \emph {et~al.}}]{Berti:2015itd}%
  \BibitemOpen
  \bibfield  {author} {\bibinfo {author} {\bibfnamefont {E.}~\bibnamefont
  {Berti}} \emph {et~al.},\ }\href
  {https://doi.org/10.1088/0264-9381/32/24/243001} {\bibfield  {journal}
  {\bibinfo  {journal} {Class. Quant. Grav.}\ }\textbf {\bibinfo {volume}
  {32}},\ \bibinfo {pages} {243001} (\bibinfo {year} {2015})},\ \Eprint
  {https://arxiv.org/abs/1501.07274} {arXiv:1501.07274 [gr-qc]} \BibitemShut
  {NoStop}%
\bibitem [{\citenamefont {Cardoso}\ and\ \citenamefont
  {Gualtieri}(2016)}]{Cardoso:2016ryw}%
  \BibitemOpen
  \bibfield  {author} {\bibinfo {author} {\bibfnamefont {V.}~\bibnamefont
  {Cardoso}}\ and\ \bibinfo {author} {\bibfnamefont {L.}~\bibnamefont
  {Gualtieri}},\ }\href {https://doi.org/10.1088/0264-9381/33/17/174001}
  {\bibfield  {journal} {\bibinfo  {journal} {Class. Quant. Grav.}\ }\textbf
  {\bibinfo {volume} {33}},\ \bibinfo {pages} {174001} (\bibinfo {year}
  {2016})},\ \Eprint {https://arxiv.org/abs/1607.03133} {arXiv:1607.03133
  [gr-qc]} \BibitemShut {NoStop}%
\bibitem [{\citenamefont {Ryan}(1997)}]{ryan_spinning_1997}%
  \BibitemOpen
  \bibfield  {author} {\bibinfo {author} {\bibfnamefont {F.~D.}\ \bibnamefont
  {Ryan}},\ }\href {https://doi.org/10.1103/PhysRevD.55.6081} {\bibfield
  {journal} {\bibinfo  {journal} {Physical Review D}\ }\textbf {\bibinfo
  {volume} {55}},\ \bibinfo {pages} {6081} (\bibinfo {year}
  {1997})}\BibitemShut {NoStop}%
\bibitem [{\citenamefont {Barack}\ and\ \citenamefont
  {Cutler}(2007)}]{Barack:2006pq}%
  \BibitemOpen
  \bibfield  {author} {\bibinfo {author} {\bibfnamefont {L.}~\bibnamefont
  {Barack}}\ and\ \bibinfo {author} {\bibfnamefont {C.}~\bibnamefont
  {Cutler}},\ }\href {https://doi.org/10.1103/PhysRevD.75.042003} {\bibfield
  {journal} {\bibinfo  {journal} {Phys. Rev. D}\ }\textbf {\bibinfo {volume}
  {75}},\ \bibinfo {pages} {042003} (\bibinfo {year} {2007})},\ \Eprint
  {https://arxiv.org/abs/gr-qc/0612029} {arXiv:gr-qc/0612029} \BibitemShut
  {NoStop}%
\bibitem [{\citenamefont {Krishnendu}\ \emph {et~al.}(2017)\citenamefont
  {Krishnendu}, \citenamefont {Arun},\ and\ \citenamefont
  {Mishra}}]{Krishnendu:2017shb}%
  \BibitemOpen
  \bibfield  {author} {\bibinfo {author} {\bibfnamefont {N.~V.}\ \bibnamefont
  {Krishnendu}}, \bibinfo {author} {\bibfnamefont {K.~G.}\ \bibnamefont
  {Arun}},\ and\ \bibinfo {author} {\bibfnamefont {C.~K.}\ \bibnamefont
  {Mishra}},\ }\href {https://doi.org/10.1103/PhysRevLett.119.091101}
  {\bibfield  {journal} {\bibinfo  {journal} {Phys. Rev. Lett.}\ }\textbf
  {\bibinfo {volume} {119}},\ \bibinfo {pages} {091101} (\bibinfo {year}
  {2017})},\ \Eprint {https://arxiv.org/abs/1701.06318} {arXiv:1701.06318
  [gr-qc]} \BibitemShut {NoStop}%
\bibitem [{\citenamefont {Krishnendu}\ and\ \citenamefont
  {Yelikar}(2020)}]{Krishnendu:2019ebd}%
  \BibitemOpen
  \bibfield  {author} {\bibinfo {author} {\bibfnamefont {N.~V.}\ \bibnamefont
  {Krishnendu}}\ and\ \bibinfo {author} {\bibfnamefont {A.~B.}\ \bibnamefont
  {Yelikar}},\ }\href {https://doi.org/10.1088/1361-6382/ababb1} {\bibfield
  {journal} {\bibinfo  {journal} {Class. Quant. Grav.}\ }\textbf {\bibinfo
  {volume} {37}},\ \bibinfo {pages} {205019} (\bibinfo {year} {2020})},\
  \Eprint {https://arxiv.org/abs/1904.12712} {arXiv:1904.12712 [gr-qc]}
  \BibitemShut {NoStop}%
\bibitem [{\citenamefont {Herdeiro}\ and\ \citenamefont
  {Radu}(2018)}]{herdeiro_spinning_2018}%
  \BibitemOpen
  \bibfield  {author} {\bibinfo {author} {\bibfnamefont {C.~A.~R.}\
  \bibnamefont {Herdeiro}}\ and\ \bibinfo {author} {\bibfnamefont
  {E.}~\bibnamefont {Radu}},\ }\href
  {https://doi.org/10.1142/S0218271818430095} {\bibfield  {journal} {\bibinfo
  {journal} {International Journal of Modern Physics D}\ }\textbf {\bibinfo
  {volume} {27}},\ \bibinfo {pages} {1843009} (\bibinfo {year}
  {2018})}\BibitemShut {NoStop}%
\bibitem [{\citenamefont {Komatsu}\ \emph {et~al.}(1989)\citenamefont
  {Komatsu}, \citenamefont {Eriguchi},\ and\ \citenamefont
  {Hachisu}}]{komatsu_rapidly_1989}%
  \BibitemOpen
  \bibfield  {author} {\bibinfo {author} {\bibfnamefont {H.}~\bibnamefont
  {Komatsu}}, \bibinfo {author} {\bibfnamefont {Y.}~\bibnamefont {Eriguchi}},\
  and\ \bibinfo {author} {\bibfnamefont {I.}~\bibnamefont {Hachisu}},\ }\href
  {https://doi.org/10.1093/mnras/237.2.355} {\bibfield  {journal} {\bibinfo
  {journal} {Monthly Notices of the Royal Astronomical Society}\ }\textbf
  {\bibinfo {volume} {237}},\ \bibinfo {pages} {355} (\bibinfo {year}
  {1989})}\BibitemShut {NoStop}%
\bibitem [{\citenamefont {Hansen}(1974{\natexlab{b}})}]{hansen_multipole_1974}%
  \BibitemOpen
  \bibfield  {author} {\bibinfo {author} {\bibfnamefont {R.~O.}\ \bibnamefont
  {Hansen}},\ }\href {https://doi.org/10.1063/1.1666501} {\bibfield  {journal}
  {\bibinfo  {journal} {Journal of Mathematical Physics}\ }\textbf {\bibinfo
  {volume} {15}},\ \bibinfo {pages} {46} (\bibinfo {year}
  {1974}{\natexlab{b}})}\BibitemShut {NoStop}%
\bibitem [{\citenamefont {Thorne}(1980)}]{thorne_multipole_1980}%
  \BibitemOpen
  \bibfield  {author} {\bibinfo {author} {\bibfnamefont {K.~S.}\ \bibnamefont
  {Thorne}},\ }\href {https://doi.org/10.1103/RevModPhys.52.299} {\bibfield
  {journal} {\bibinfo  {journal} {Reviews of Modern Physics}\ }\textbf
  {\bibinfo {volume} {52}},\ \bibinfo {pages} {299} (\bibinfo {year}
  {1980})}\BibitemShut {NoStop}%
\bibitem [{\citenamefont {Pappas}\ and\ \citenamefont
  {Apostolatos}(2012)}]{pappas_revising_2012}%
  \BibitemOpen
  \bibfield  {author} {\bibinfo {author} {\bibfnamefont {G.}~\bibnamefont
  {Pappas}}\ and\ \bibinfo {author} {\bibfnamefont {T.~A.}\ \bibnamefont
  {Apostolatos}},\ }\href {https://doi.org/10.1103/PhysRevLett.108.231104}
  {\bibfield  {journal} {\bibinfo  {journal} {Physical Review Letters}\
  }\textbf {\bibinfo {volume} {108}},\ \bibinfo {pages} {231104} (\bibinfo
  {year} {2012})}\BibitemShut {NoStop}%
\bibitem [{\citenamefont {Urbanec}\ \emph {et~al.}(2013)\citenamefont
  {Urbanec}, \citenamefont {Miller},\ and\ \citenamefont
  {Stuchlik}}]{Urbanec:2013fs}%
  \BibitemOpen
  \bibfield  {author} {\bibinfo {author} {\bibfnamefont {M.}~\bibnamefont
  {Urbanec}}, \bibinfo {author} {\bibfnamefont {J.~C.}\ \bibnamefont
  {Miller}},\ and\ \bibinfo {author} {\bibfnamefont {Z.}~\bibnamefont
  {Stuchlik}},\ }\href {https://doi.org/10.1093/mnras/stt858} {\bibfield
  {journal} {\bibinfo  {journal} {Mon. Not. Roy. Astron. Soc.}\ }\textbf
  {\bibinfo {volume} {433}},\ \bibinfo {pages} {1903} (\bibinfo {year}
  {2013})},\ \Eprint {https://arxiv.org/abs/1301.5925} {arXiv:1301.5925
  [astro-ph.SR]} \BibitemShut {NoStop}%
\bibitem [{\citenamefont {Yagi}\ and\ \citenamefont
  {Yunes}(2017)}]{Yagi:2016bkt}%
  \BibitemOpen
  \bibfield  {author} {\bibinfo {author} {\bibfnamefont {K.}~\bibnamefont
  {Yagi}}\ and\ \bibinfo {author} {\bibfnamefont {N.}~\bibnamefont {Yunes}},\
  }\href {https://doi.org/10.1016/j.physrep.2017.03.002} {\bibfield  {journal}
  {\bibinfo  {journal} {Phys. Rept.}\ }\textbf {\bibinfo {volume} {681}},\
  \bibinfo {pages} {1} (\bibinfo {year} {2017})},\ \Eprint
  {https://arxiv.org/abs/1608.02582} {arXiv:1608.02582 [gr-qc]} \BibitemShut
  {NoStop}%
\bibitem [{web()}]{webpage}%
  \BibitemOpen
  \href@noop {} {\ }\bibinfo {note}
  {\noindent\href{https://web.uniroma1.it/gmunu/}{https://web.uniroma1.it/gmunu/}}\BibitemShut
  {NoStop}%
\bibitem [{\citenamefont {Poisson}(2004)}]{poisson_relativists_2004}%
  \BibitemOpen
  \bibfield  {author} {\bibinfo {author} {\bibfnamefont {E.}~\bibnamefont
  {Poisson}},\ }\href {https://doi.org/10.1017/CBO9780511606601} {\emph
  {\bibinfo {title} {A Relativist's Toolkit: The Mathematics of Black-Hole
  Mechanics}}}\ (\bibinfo  {publisher} {Cambridge University Press},\ \bibinfo
  {address} {Cambridge},\ \bibinfo {year} {2004})\BibitemShut {NoStop}%
\bibitem [{\citenamefont {Herdeiro}\ \emph {et~al.}(2015)\citenamefont
  {Herdeiro}, \citenamefont {Radu},\ and\ \citenamefont
  {Rúnarsson}}]{herdeiro_kerr_2015}%
  \BibitemOpen
  \bibfield  {author} {\bibinfo {author} {\bibfnamefont {C.~A.}\ \bibnamefont
  {Herdeiro}}, \bibinfo {author} {\bibfnamefont {E.}~\bibnamefont {Radu}},\
  and\ \bibinfo {author} {\bibfnamefont {H.}~\bibnamefont {Rúnarsson}},\
  }\href {https://doi.org/10.1103/PhysRevD.92.084059} {\bibfield  {journal}
  {\bibinfo  {journal} {Physical Review D}\ }\textbf {\bibinfo {volume} {92}},\
  \bibinfo {pages} {084059} (\bibinfo {year} {2015})}\BibitemShut {NoStop}%
\bibitem [{\citenamefont {Delgado}\ \emph {et~al.}(2020)\citenamefont
  {Delgado}, \citenamefont {Herdeiro},\ and\ \citenamefont
  {Radu}}]{delgado_rotating_2020}%
  \BibitemOpen
  \bibfield  {author} {\bibinfo {author} {\bibfnamefont {J.~F.}\ \bibnamefont
  {Delgado}}, \bibinfo {author} {\bibfnamefont {C.~A.}\ \bibnamefont
  {Herdeiro}},\ and\ \bibinfo {author} {\bibfnamefont {E.}~\bibnamefont
  {Radu}},\ }\href {https://doi.org/10.1088/1475-7516/2020/06/037} {\bibfield
  {journal} {\bibinfo  {journal} {Journal of Cosmology and Astroparticle
  Physics}\ }\textbf {\bibinfo {volume} {2020}}\bibinfo  {number} { (06)},\
  \bibinfo {pages} {037}}\BibitemShut {NoStop}%
\bibitem [{\citenamefont {Buchdahl}(1959)}]{Buchdahl:1959zz}%
  \BibitemOpen
\bibfield  {number} {  }\bibfield  {author} {\bibinfo {author} {\bibfnamefont
  {H.~A.}\ \bibnamefont {Buchdahl}},\ }\href
  {https://doi.org/10.1103/PhysRev.116.1027} {\bibfield  {journal} {\bibinfo
  {journal} {Phys. Rev.}\ }\textbf {\bibinfo {volume} {116}},\ \bibinfo {pages}
  {1027} (\bibinfo {year} {1959})}\BibitemShut {NoStop}%
\bibitem [{\citenamefont {{Friedman}}(1978)}]{1978CMaPh..63..243F}%
  \BibitemOpen
  \bibfield  {author} {\bibinfo {author} {\bibfnamefont {J.~L.}\ \bibnamefont
  {{Friedman}}},\ }\href {https://doi.org/10.1007/BF01196933} {\bibfield
  {journal} {\bibinfo  {journal} {Communications in Mathematical Physics}\
  }\textbf {\bibinfo {volume} {63}},\ \bibinfo {pages} {243} (\bibinfo {year}
  {1978})}\BibitemShut {NoStop}%
\bibitem [{\citenamefont {Cardoso}\ \emph {et~al.}(2008)\citenamefont
  {Cardoso}, \citenamefont {Pani}, \citenamefont {Cadoni},\ and\ \citenamefont
  {Cavaglia}}]{Cardoso:2007az}%
  \BibitemOpen
  \bibfield  {author} {\bibinfo {author} {\bibfnamefont {V.}~\bibnamefont
  {Cardoso}}, \bibinfo {author} {\bibfnamefont {P.}~\bibnamefont {Pani}},
  \bibinfo {author} {\bibfnamefont {M.}~\bibnamefont {Cadoni}},\ and\ \bibinfo
  {author} {\bibfnamefont {M.}~\bibnamefont {Cavaglia}},\ }\href
  {https://doi.org/10.1103/PhysRevD.77.124044} {\bibfield  {journal} {\bibinfo
  {journal} {Phys.Rev.}\ }\textbf {\bibinfo {volume} {D77}},\ \bibinfo {pages}
  {124044} (\bibinfo {year} {2008})},\ \Eprint
  {https://arxiv.org/abs/0709.0532} {arXiv:0709.0532 [gr-qc]} \BibitemShut
  {NoStop}%
\bibitem [{\citenamefont {Yagi}\ and\ \citenamefont
  {Yunes}(2013)}]{yagi_i-love-q_2013}%
  \BibitemOpen
  \bibfield  {author} {\bibinfo {author} {\bibfnamefont {K.}~\bibnamefont
  {Yagi}}\ and\ \bibinfo {author} {\bibfnamefont {N.}~\bibnamefont {Yunes}},\
  }\href {https://doi.org/10.1103/PhysRevD.88.023009} {\bibfield  {journal}
  {\bibinfo  {journal} {Physical Review D}\ }\textbf {\bibinfo {volume} {88}},\
  \bibinfo {pages} {023009} (\bibinfo {year} {2013})},\ \bibinfo {note} {arXiv:
  1303.1528}\BibitemShut {NoStop}%
\bibitem [{\citenamefont {Sennett}\ \emph {et~al.}(2017)\citenamefont
  {Sennett}, \citenamefont {Hinderer}, \citenamefont {Steinhoff}, \citenamefont
  {Buonanno},\ and\ \citenamefont {Ossokine}}]{Sennett:2017etc}%
  \BibitemOpen
  \bibfield  {author} {\bibinfo {author} {\bibfnamefont {N.}~\bibnamefont
  {Sennett}}, \bibinfo {author} {\bibfnamefont {T.}~\bibnamefont {Hinderer}},
  \bibinfo {author} {\bibfnamefont {J.}~\bibnamefont {Steinhoff}}, \bibinfo
  {author} {\bibfnamefont {A.}~\bibnamefont {Buonanno}},\ and\ \bibinfo
  {author} {\bibfnamefont {S.}~\bibnamefont {Ossokine}},\ }\href
  {https://doi.org/10.1103/PhysRevD.96.024002} {\bibfield  {journal} {\bibinfo
  {journal} {Phys. Rev. D}\ }\textbf {\bibinfo {volume} {96}},\ \bibinfo
  {pages} {024002} (\bibinfo {year} {2017})},\ \Eprint
  {https://arxiv.org/abs/1704.08651} {arXiv:1704.08651 [gr-qc]} \BibitemShut
  {NoStop}%
\bibitem [{\citenamefont {Herdeiro}\ and\ \citenamefont
  {Radu}(2014)}]{Herdeiro:2014goa}%
  \BibitemOpen
  \bibfield  {author} {\bibinfo {author} {\bibfnamefont {C.~A.~R.}\
  \bibnamefont {Herdeiro}}\ and\ \bibinfo {author} {\bibfnamefont
  {E.}~\bibnamefont {Radu}},\ }\href
  {https://doi.org/10.1103/PhysRevLett.112.221101} {\bibfield  {journal}
  {\bibinfo  {journal} {Phys. Rev. Lett.}\ }\textbf {\bibinfo {volume} {112}},\
  \bibinfo {pages} {221101} (\bibinfo {year} {2014})}\BibitemShut {NoStop}%
\bibitem [{\citenamefont {Herdeiro}\ and\ \citenamefont
  {Radu}(2015{\natexlab{a}})}]{Herdeiro:2015waa}%
  \BibitemOpen
  \bibfield  {author} {\bibinfo {author} {\bibfnamefont {C.~A.~R.}\
  \bibnamefont {Herdeiro}}\ and\ \bibinfo {author} {\bibfnamefont
  {E.}~\bibnamefont {Radu}},\ }\href
  {https://doi.org/10.1142/S0218271815420146} {\bibfield  {journal} {\bibinfo
  {journal} {Int. J. Mod. Phys. D}\ }\textbf {\bibinfo {volume} {24}},\
  \bibinfo {pages} {1542014} (\bibinfo {year} {2015}{\natexlab{a}})},\ \Eprint
  {https://arxiv.org/abs/1504.08209} {arXiv:1504.08209 [gr-qc]} \BibitemShut
  {NoStop}%
\bibitem [{\citenamefont {Herdeiro}\ and\ \citenamefont
  {Radu}(2015{\natexlab{b}})}]{Herdeiro:2015gia}%
  \BibitemOpen
  \bibfield  {author} {\bibinfo {author} {\bibfnamefont {C.}~\bibnamefont
  {Herdeiro}}\ and\ \bibinfo {author} {\bibfnamefont {E.}~\bibnamefont
  {Radu}},\ }\href {https://doi.org/10.1088/0264-9381/32/14/144001} {\bibfield
  {journal} {\bibinfo  {journal} {Class. Quant. Grav.}\ }\textbf {\bibinfo
  {volume} {32}},\ \bibinfo {pages} {144001} (\bibinfo {year}
  {2015}{\natexlab{b}})},\ \Eprint {https://arxiv.org/abs/1501.04319}
  {arXiv:1501.04319 [gr-qc]} \BibitemShut {NoStop}%
\bibitem [{\citenamefont {Herdeiro}\ \emph {et~al.}(2016)\citenamefont
  {Herdeiro}, \citenamefont {Radu},\ and\ \citenamefont
  {R\'unarsson}}]{Herdeiro:2016tmi}%
  \BibitemOpen
  \bibfield  {author} {\bibinfo {author} {\bibfnamefont {C.}~\bibnamefont
  {Herdeiro}}, \bibinfo {author} {\bibfnamefont {E.}~\bibnamefont {Radu}},\
  and\ \bibinfo {author} {\bibfnamefont {H.}~\bibnamefont {R\'unarsson}},\
  }\href {https://doi.org/10.1088/0264-9381/33/15/154001} {\bibfield  {journal}
  {\bibinfo  {journal} {Class. Quant. Grav.}\ }\textbf {\bibinfo {volume}
  {33}},\ \bibinfo {pages} {154001} (\bibinfo {year} {2016})},\ \Eprint
  {https://arxiv.org/abs/1603.02687} {arXiv:1603.02687 [gr-qc]} \BibitemShut
  {NoStop}%
\end{thebibliography}%

\end{document}